\documentclass[useAMS,usenatbib]{mn2e}

\usepackage{graphicx}
\usepackage{amssymb}
\usepackage{amsmath}
\usepackage{datetime}
\usepackage[normalem]{ulem}
\usepackage{times}
\newcommand{\apjs}{ApJS}
\newcommand{\apjl}{ApJ}

\newcommand{\aap}{A\&A}

\usepackage{color}
\usepackage{soul}
\definecolor{postref}{rgb}{1,0,0}

\date{\currenttime \today}

\begin{document}

\title[Effects of Cooling on X-rays]
{
X-rays from Magnetically Confined Wind Shocks:\\
Effect of Cooling-Regulated Shock Retreat
}

 \author[A. ud-Doula et al.]
{Asif ud-Doula$^1$\thanks{Email: asif@psu.edu},
 Stanley Owocki$^2$,
 Richard Townsend$^3$,
 Veronique Petit$^2$
 \newauthor
\&   David Cohen$^4$ \\
 $^1$ Penn State Worthington Scranton, Dunmore, PA 18512, USA.\\
 $^2$ Department of Physics and Astronomy, Bartol Research Institute,
 University of Delaware, Newark, DE 19716, USA\\
 $^3$ Department of Astronomy, University of Wisconsin-Madison, 5534 Sterling Hall, 475 N Charter Street, Madison, WI 53706, USA \\
 $^4$ Department of Physics and Astronomy, Swarthmore College, Swarthmore, PA 19081, USA
}

\def\<<{{\ll}}
\def\>>{{\gg}}
\def\wig{{\sim}}
\def\spose#1{\hbox to 0pt{#1\hss}}
\def\ltwig{\mathrel{\spose{\lower 3pt\hbox{$\mathchar"218$}}
     R_{\rm A}ise 2.0pt\hbox{$\mathchar"13C$}}}
\def\gtwig{\mathrel{\spose{\lower 3pt\hbox{$\mathchar"218$}}
     R_{\rm A}ise 2.0pt\hbox{$\mathchar"13E$}}}
\def\+/-{{\pm}}
\def\=={{\equiv}}
\def\mubar{{\bar \mu}}
\def\mustar{\mu_{\ast}}
\def\Lambar{{\bar \Lambda}}
\def\Rstar{R_{\ast}}
\def\Mstar{M_{\ast}}
\def\Lstar{L_{\ast}}
\def\Tstar{T_{\ast}}
\def\gstar{g_{\ast}}
\def\vth{v_{th}}
\def\grad{g_{rad}}
\def\glines{g_{lines}}
\def\Mdot{\dot M}
\def\mdot{\dot m}
\def\yr{{\rm yr}}
\def\ksec{{\rm ksec}}
\def\kms{{\rm km/s}}
\def\qad{\dot q_{ad}}
\def\qlines{\dot q_{lines}}
\def\solar{\odot}
\def\Msun{M_{\solar}}
\def\msbyr{\Msun/\yr}
\def\Rsun{R_{\solar}}
\def\Lsun{L_{\solar}}
\def\Be{{\rm Be}}
\def\Rpole{R_{p}}
\def\Req{R_{eq}}
\def\Rmin{R_{min}}
\def\Rmax{R_{max}}
\def\Rstag{R_{stag}}
\def\vinf{V_\infty}
\def\Vrot{V_{rot}}
\def\Vcrit{V_{crit}}
\def\half{{1 \over 2}}
\newcommand{\beq}{\begin{equation}}
\newcommand{\eeq}{\end{equation}}
\newcommand{\beqa}{\begin{eqnarray}}
\newcommand{\eeqa}{\end{eqnarray}}
\def\phip{{\phi'}}

\maketitle

\begin{abstract}
We use 2D MHD simulations to examine the effects of radiative cooling and inverse Compton (IC) cooling on X-ray emission from magnetically confined wind shocks (MCWS) in magnetic massive stars with radiatively driven stellar winds.
For the standard dependence of mass loss rate on luminosity $\Mdot \sim L^{1.7} $, 
the scaling of IC cooling with $L$ and radiative cooling with $\Mdot$ means that IC cooling become formally more important for lower luminosity stars. 
However, because the sense of the trends is similar, we find the overall effect of including IC cooling is quite modest.
More significantly, for stars with high enough mass loss to keep the shocks radiative, the MHD simulations indicate a linear scaling of X-ray luminosity with mass loss rate;
but for lower luminosity stars with weak winds, X-ray emission is reduced and softened by
a {\em shock retreat} resulting from the larger post-shock cooling length, which within the fixed length of a closed magnetic loop forces the shock back to lower pre-shock wind speeds.
A semi-analytic scaling analysis that accounts both for the wind magnetic confinement and this shock retreat yields  X-ray luminosities that 
have a similar scaling trend, but a factor few higher values, compared to
time-averages computed from the MHD simulations.
The simulation and scaling results here thus provide a good basis for interpreting available X-ray observations from the growing list of massive stars with confirmed large-scale magnetic fields.
\end{abstract}


\section{INTRODUCTION}

Hot luminous, massive stars of spectral type O and B are prominent sources of X-rays thought to originate from shocks in their high-speed, radiatively driven stellar winds.
In putatively single, non-magnetic O stars, the intrinsic instability of wind driving by line-scattering 
leads to embedded wind shocks that are thought to be the source of  their relatively soft X-rays ($\sim$0.5\,keV) X-ray spectrum, with a total X-ray luminosity that scales with stellar bolometric luminosity, $L_{\rm x} \sim L_{\rm bol}$
 \citep{Chlebowski89,   
 Naze11, Owocki13}.
In  massive binary systems the collision of the two stellar winds at up to the wind terminal speeds can lead to even higher $L_x$, generally with a significantly harder (up to 10\,keV)  spectrum \citep{Stevens92, Gagne11}.

The study here examines a third source of X-rays from OB winds, namely those observed from the subset ($\sim$10\%) of massive stars with strong, globally ordered (often significantly dipolar) magnetic fields
\citep{Petit13}; 
in this case, the trapping and channeling of the stellar wind in closed magnetic loops leads to {\em magnetically confined wind shocks}  (MCWS) \citep[hereafter BM97a,b]{Babel97,Babel97b}, with pre-shock flow speeds that are some fraction of the wind  terminal speed, resulting in intermediate energies for the shocks and associated X-rays  ($\sim$2\,keV).
A prototypical example is provided by the magnetic O-type  star $\theta^1$~Ori~C, which shows moderately hard X-ray emission with a rotational phase variation that matches well the expectations of the MCWS paradigm 
\citep{Gagne05}.

Our approach here builds on our previous MHD simulation studies of the role of magnetic fields in wind channeling 
 \citep[Paper I]{Uddoula02},
including its combined effect with stellar rotation in formation of centrifugally supported magnetospheres  
 \citep[Paper II] {Uddoula08} 
and in enhancing the angular momentum loss from the stellar wind 
\citep[Paper III]{Uddoula09}.
In contrast to the assumption of isothermal flow used in these studies, our examination here of X-ray emission now requires a full treatment of the wind energy balance, including the cooling of shock-heated gas.
This follows our successful specific application of MHD simulations of MCWS with a full energy balance for modeling X-ray observations of 
$\theta^1$~Ori~C 
\citep{Gagne05}.
But rather than focus on any specific star, the aim here is to derive broad scaling relations for how the X-ray luminosity and spectral properties depend on the stellar luminosity $L$ and mass loss rate $\Mdot$, with particular attention to how these affect the efficiency of shock cooling.
The initial study here will neglect rotation, and so focus on stars with ``dynamical magnetospheres'' (DM), deferring to future work studies of the effect of rapid rotation on X-rays from ``centrifugal magnetospheres'' (CM) \citep{Sundqvist12c, Petit13}.

For high-density winds with efficient shock cooling, the maximum shock strength depends on  the  speed reached before the flow from opposite footpoints of a closed loop collide near the loop top, and thus on the maximum loop height. 
The analyses in papers I-III show that this is generally somewhat below [see eqn.\ (\ref{eq:rcetas})] the characteristic wind Alfv\'{e}n radius $R_A$, which for a dipole field scales as a factor $\sim \eta_\ast^{1/4}$  times the stellar radius $\Rstar$, where 
\beq
\eta_\ast \equiv  \frac{B_{eq}^2 \Rstar^2}{\Mdot \vinf}
\label{eq:etasdef}
\eeq
 is the ``wind magnetic confinement parameter'' for an equatorial surface field $B_{eq}$, with $\Mdot$ and $\vinf$ the wind mass loss rate and terminal speed that would occur in {\em non-magnetic} star with the same stellar parameters.
For magnetic O-stars with $\eta_\ast \approx 10-100$, the associated Alfv\'{e}n radii $R_A \approx 1.7 - 3 \Rstar$ allow acceleration up to half terminal speed, typically about 1500~km/s. This leads to shock energies $\sim$2\,keV that are sufficient to explain the moderately hard X-rays observed in $\theta^1$~Ori~C \citep{Gagne05}.

For magnetic B-type stars, the combination of lower mass loss rates ($\Mdot < 10^{-9} \Msun$/yr) and  very strong (1-10\,kG) fields leads to very strong magnetic confinement,  with $\eta_\ast \sim 10^4 - 10^6$ and so much larger
Alfv\'{e}n radii, $R_A \approx 10-30 \Rstar$. This would suggest a potential to accelerate the flow to near the wind terminal speed $\sim$\, 3000 km\,s$^{-1}$ within closed magnetic loops, and so yield much stronger shocks (up to 10\,keV) and thus much harder X-rays.

\begin{figure}
\begin{center}
\vfill
\includegraphics[scale=0.65]{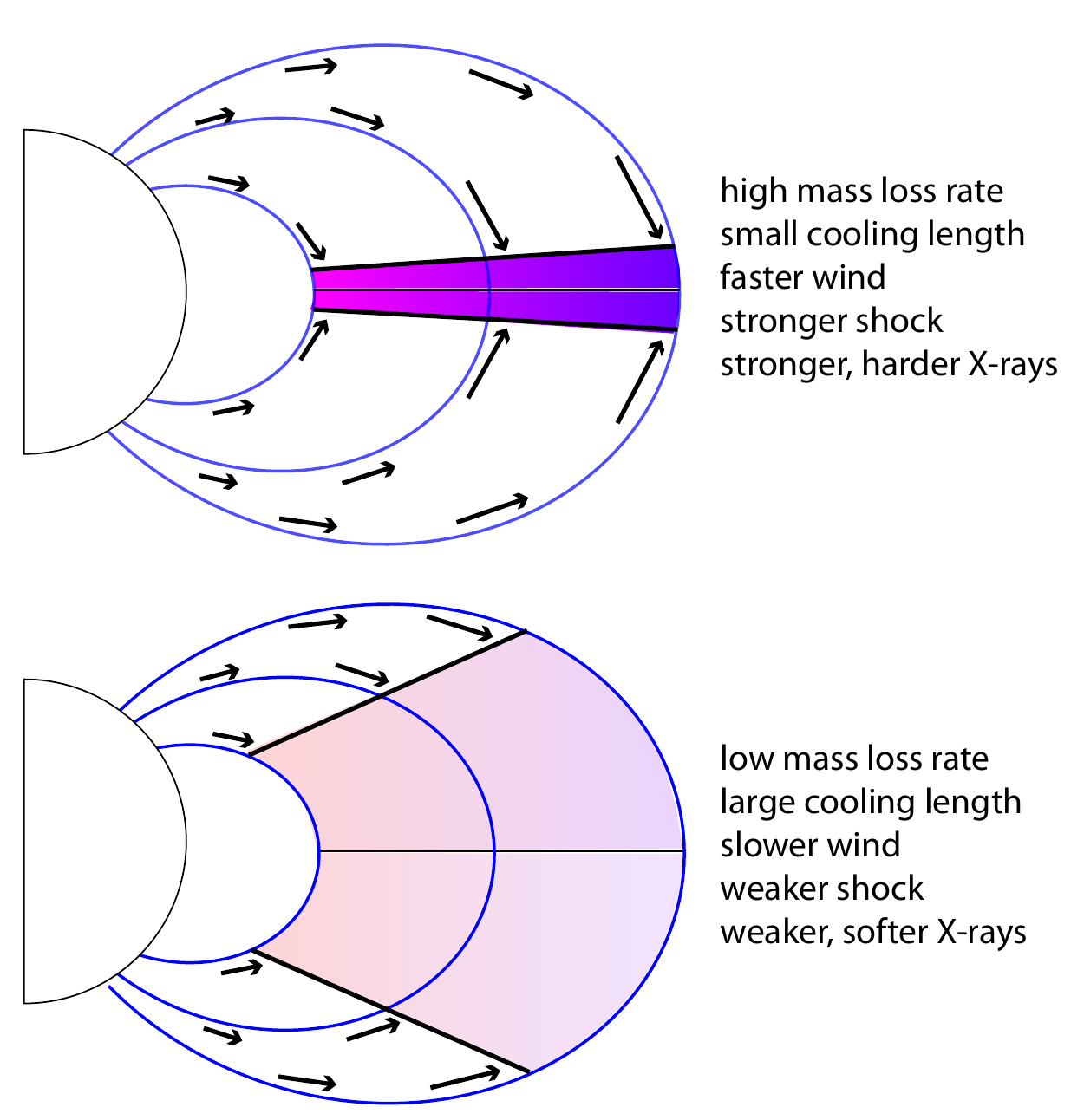}
\caption
{Schematic illustration of the ``shock retreat'' from inefficient cooling associated with a lower mass loss rate ${\dot M}$, showing a hemispheric, planar slice of a stellar dipole magnetic field.
Wind outflow driven from opposite foot-points of closed magnetic loops is channeled into a collision near the loop top, forming magnetically confined wind shocks (MCWS).
For the high ${\dot M}$ case in the upper panel, the efficient cooling keeps the shock-heated gas within a narrow cooling layer, allowing the pre-shock wind to accelerate to a high speed and so produce strong shocks with strong, relatively hard X-ray emission.
For the low ${\dot M}$ case in the lower panel, the inefficient cooling forces a shock retreat down to lower radii with slower pre-shock wind,  leading to weaker shocks with weaker, softer X-ray emission.
}
\label{fig:shockretreat}
\end{center}
\end{figure}

However, as illustrated schematically in figure \ref{fig:shockretreat}  (see also figure 13 of BM97a) and quantified further below, the much lower mass loss rates of such B-stars also implies much less efficient cooling of the post-shock flow. When the associated cooling length becomes comparable to the Alfv\'{e}n radius, the shock location is effectively forced to ``retreat'' back down the loop, to a lower radius where the lower wind speed yields a weaker shock, implying then a much softer X-ray spectrum. 

To quantify this  {\em shock retreat} effect, and derive general scalings for how the X-ray luminosity and hardness depend on the stellar luminosity and associated wind mass loss rate,  the analysis here carries out an extensive parameter study based on 2D MHD simulations with a detailed energy balance.
To focus on the relative roles of magnetic confinement and shock cooling, we ignore here the effects of stellar rotation, since this would introduce a third free parameter to our variations of magnetic confinement  and  cooling efficiency.

As a prelude to the detailed MHD simulation study in \S\S\,3-4, the next section (\S 2) develops the basic equations, and presents an analysis of the relative importance of both radiative and inverse Compton (IC) cooling in stars of various luminosities and mass loss rates.
In \S 3, the full 2D MHD simulation results (for a standard model appropriate to O-type supergiant star with large mass loss rate and so strong radiative cooling) are used to derive differential emission measure (DEM) and associated dynamic X-ray spectra.
\S\,4  then presents a general parameter study for how the X-ray emission in this standard model scales with a modified cooling efficiency, intended as a proxy for varying the wind mass loss rate.
Comparisons with a semi-analytic scaling analysis (\S 4.4) indicate that X-ray luminosity depends on both the magnetic confinement parameters $\eta_\ast$ and a radiative cooling parameter $\chi_\infty$ [see eqn.\ (\ref{eq:chiinf})], providing then a generalized scaling law [eqn.\ (\ref{eq:lxlkin})] for interpreting X-ray observations for magnetic massive stars with a range of stellar parameters.
The concluding section (\S 5) summarizes results and their implications for interpreting X-ray observations, and outlines directions of future work.

\section{ENERGY BALANCE IN WIND SHOCKS}

\subsection{MHD equations}

As in papers I-III, our general approach is to use the {\sc ZEUS-3D}
\citep{Stone92} numerical magneto-hydrodynamics (MHD) code  to evolve a 2D consistent dynamical
solution for a line-driven stellar wind
from a non-rotating star with a dipole surface field.
In vector form, the MHD treatment includes equations for mass continuity,
\beq
\frac{D\rho}{Dt} 
 + \rho \nabla \cdot {\bf v} = 0 \, ,
\label{eq:masscon}
\eeq
and momentum balance,
\beq
\frac{D{\bf v}}{Dt} =
- \frac{\nabla p}{\rho} +\frac{1}{4 \pi \rho} (\nabla \times
{\bf B}) \times {\bf B} - { GM
{\hat {\bf r}} \over r^2 }
+ {\bf g}_{\rm lines}
,
\label{eq:eom}
\eeq
where
$D/Dt = \partial/\partial t + {\bf v} \cdot \nabla$ is the total time derivative advecting along the flow speed ${\bf v}$,
and the other notation follows common conventions,
as defined in detail in section 2 of Paper I.
(Note that eqn. (\ref{eq:eom}) here corrects some minor errors in the
corresponding eqn. (2) of Paper I.)

As in all our previous MHD studies,
the treatment of the acceleration ${\bf g}_{\rm lines}$ by line-scattering follows the standard  \citet*[hereafter CAK]{Castor75} formalism,
corrected for the finite cone angle of the star, using a spherical expansion approximation for the local flow gradients
\citep*{Pauldrach86, Friend86}, and ignoring  {\em non-radial} components of the line-force.

By the ideal gas law, the pressure, density and temperature are related through $p=kT/\mubar$, where $k$ is Boltzmann's constant, and the mean molecular weight $\mubar \approx 0.62 \, m_p$, with $m_p$ the proton mass.

\subsection{Energy balance}
\label{sec:enbal}
Instead of the {\em isothermal} approximation used in Papers I-III,
 we now include a full energy equation.
For a monatomic ideal gas with ratio of specific heats $\gamma=5/3$,  the internal energy density is related to the pressure by $e=p/(\gamma -1) = (3/2) p$. 
In analogy with the mass conservation (\ref{eq:masscon}),
the energy balance can be written in a conservation form, but now with non-zero terms on the right-hand-side to account for the sources and sinks of energy,
\beq
\frac{\partial e}{\partial t} + \nabla \cdot (e {\bf v}) =
- p \nabla \cdot {\bf v}
+ Q - C
 \, .
 \label{eq:encon}
\eeq
Here the pressure term represents the effect of compressive heating ($\nabla \cdot {\bf v} < 0$) or expansive cooling ($\nabla \cdot {\bf v} >  0$), and the $Q-C$ terms account for additional volumetric heating or cooling effects.
In hot-star winds, UV photoionization heating sets a floor to the wind temperature on the order the stellar effective temperature \citep{Drew89}, but otherwise such heating is unimportant in the shock-heated regions that are the focus of the study here.
For cooling, we include here both optically thin radiative emission as well as {\em inverse Compton} (IC) cooling from scattering of the stellar UV photons by electrons that can be heated to keV energies in shocks.

For the analysis below, it is convenient to use the mass conservation (\ref{eq:masscon}) to rewrite the left side of the energy conservation (\ref{eq:encon}) in terms of the total advective time derivative of the energy per unit mass $e/\rho$, 
\beq
\rho \frac{D(e/\rho )}{Dt} = 
- p \nabla \cdot {\bf v}
 - C_{rad} - C_{\rm IC}
 \, .
 \label{eq:energy}
\eeq
The volume cooling rate from radiative emission has the scaling,
\beq
C_{rad} = n_e n_p \Lambda (T)  =   
\rho^2 \Lambda_m (T)
\, ,
\label{eq:qrad}
\eeq
where $\Lambda (T)$ is the optically thin cooling function \citep{MacDonald81, Schure09}, and the latter equality defines a mass-weighted form $\Lambda_m \equiv \Lambda/\mu_e \mu_p$.  
For a fully ionized plasma the proton and electron number densities $n_p$ and $n_e$ are related to the mass density $\rho$ through the associated hydrogen mass fraction $X= m_p/\mu_p = m_p n_p/\rho  $ and mean mass per electron  $\mu_e = \rho/n_e = 2 m_p/ (1+ X)$. We assume here the standard solar hydrogen abundance $X=0.72$.

The  IC volume cooling rate \citep{White95}
scales with the electron pressure $n_e kT = (\mubar/\mu_e) p $ and the photon energy density $U_{ph}$, 
\beq
C_{\rm IC}
 = 4 \sigma_e n_e kT  \frac{U_{ph}}{m_e c} 
 =  4  \frac{\kappa_e \mubar }{m_e c} \,  p \,  U_{ph }
\, ,
\label{eq:qic}
\eeq
where $m_e$  and $c$ are the electron mass and speed of light, and $\sigma_e$ and $\kappa_e \equiv \sigma_e/\mu_e$ are the electron scattering cross section and the associated opacity.

\subsection{Characteristic time scales}

Let us examine the time scales for the various processes in the energy equation (\ref{eq:energy}). Dividing by  the internal energy $e$, we can recast this energy equation in terms of processes leading to a change in temperature,
\beqa
- \frac{1}{T} \frac{DT}{Dt} &=&  
\frac{2}{3} \nabla \cdot {\bf v} +  
\frac{2 \mubar }{3 k} \, \frac{ \rho \Lambda_m (T)}{T} +
\frac{8}{3}  \frac{\kappa_e \mubar }{m_e c}  \,  U_{ph }
\label{eq:dTdt}
\\
- \frac{1}{t_T} &=&  ~  \frac{1}{t_{ad}} ~  + ~ \frac{1}{t_{rad}} ~ + ~ \frac{1}{t_{\rm IC}} 
\label{eq:tscales}
\, .
\eeqa
The first term on the right-hand side of (\ref{eq:dTdt}) represents the effects of heating by adiabatic compression (if $\nabla \cdot {\bf v} < 0$) or cooling by adiabatic expansion (if $\nabla \cdot {\bf v} > 0$). For the discontinuous compression at a shock, this term leads to the sudden jump in post-shock temperature. But in a wind expansion, it nominally has a cooling effect, including in the regions of a post-shock flow.
For such post-shock cooling layers, equation (\ref{eq:tscales}) thus identifies the  timescale for change in temperature with associated cooling timescales for adiabatic expansion, radiative emission, and inverse Compton scattering.

\subsection{Cooling times for a standing  shock}
\label{sec:sphshock}

As a basis for estimating the relative importance of these processes for MCWS, let us examine the scalings of the associated timescales for the simplified case of a steady, standing shock at a fixed radius $r_s$ in a steady spherical wind with specified mass loss rate $\Mdot$ and pre-shock wind speed $V_w$
\citep[see][]{Owocki13}.

For a strong shock, the immediate post-shock density is a factor 4 times the pre-shock wind value, $\rho_s = 4 \rho_w = 4 \Mdot/(4 \pi r_s^2 V_w)$.
Since the post-shock flow speed is correspondingly {\em reduced} by this factor 4, the net shock jump is $\Delta v = (3/4) V_w$, yielding a post-shock temperature
\beq
T_{\rm s} = \frac{3}{16} \frac{\mubar V_w^2}{k} \approx 14 \, {\rm MK} ~ V_8^2 
\approx 1.2 \, {\rm keV} ~ V_8^2 
\, ,
\label{eq:tsdef}
\eeq
where $V_8 \equiv V_w/(10^8$cm/s).
If we take the post-shock speed to be roughly constant and assume, for simplicity, spherical expansion
$\nabla \cdot {\bf v} = 2 v/r = V_w/2 r_s$,  then we obtain for the adiabatic expansion timescale,
\beq
t_{ad} =  3 \frac{r_s}{V_w} =
  30 \, {\rm ks}  ~ \frac{r_{12}}{V_8}
\, ,
\label{eq:texp}
\eeq
where $r_{12} = r_s/10^{12}$\,cm.

We can write the radiative cooling time as
\beq
t_{rad} 
= \frac{3 \pi k  }{2 \mubar} \, \frac{r_s^2 V_w T_s} { \Mdot  \Lambda_m (T_s) }
\approx 0.75\,{\rm ks} \, \frac{V_8^3 \, r_{12}^2}{\Mdot_{-6}}
\, ,
\label{eq:trad}
\eeq
where $\Mdot_{-6} = \Mdot/(10^{-6} \Msun/$yr) and the numerical evaluation assumes a constant cooling function, $\Lambda (T_{\rm s}) \approx  4.4 \times 10^{-23} ~ {\rm erg \, cm^3 s^{-1}}$, over the relevant range of shock temperatures, $10^{6.5} \, K < T_{\rm s} <  10^{7.5} \, K$
\citep{Schure09}.
This allows us to define a radiative vs. adiabatic cooling parameter,
\beq
\chi_{rad} \equiv \frac{t_{rad}}{t_{ad}} = 0.025  \, \frac{V_8^4 \, r_{12} }{\Mdot_{-6}}
\, .
\label{eq:chirad}
\eeq
Note that this is 0.25 times the cooling parameter defined by \citet{Stevens92} in the context of colliding stellar winds.

For stellar luminosity $L$, the photon energy density at shock radius $r_s$ is
\beq
U_{ph} = \frac{L}{4 \pi r_s^2 c} ~ \frac{2}{1+\mustar}
\, ,
\label{eq:uph}
\eeq
where the factor with $\mustar = \sqrt{1-(\Rstar/r_s)^2}$ corrects for the difference between energy density and flux for a star of radius $\Rstar$ with uniform surface brightness (i.e., ignoring limb darkening).
We then find for the IC cooling time,
\beq
t_{\rm IC} = \frac{3 \pi m_e c^2  }{2 \kappa_e \mubar} \, \frac{r_s^2 (1 + \mustar )}{L}
\approx 2.8 \, {\rm ks} \, \frac{r_{12}^2}{L_6}
\, ,
\label{eq:tic}
\eeq
where $L_6 \equiv L/(10^6 L_\odot$) and the latter approximation ignores the factor 2 variation from the $1 + 
\mustar$ term.

\begin{figure}
\begin{center}
\vfill
\includegraphics[scale=0.48]{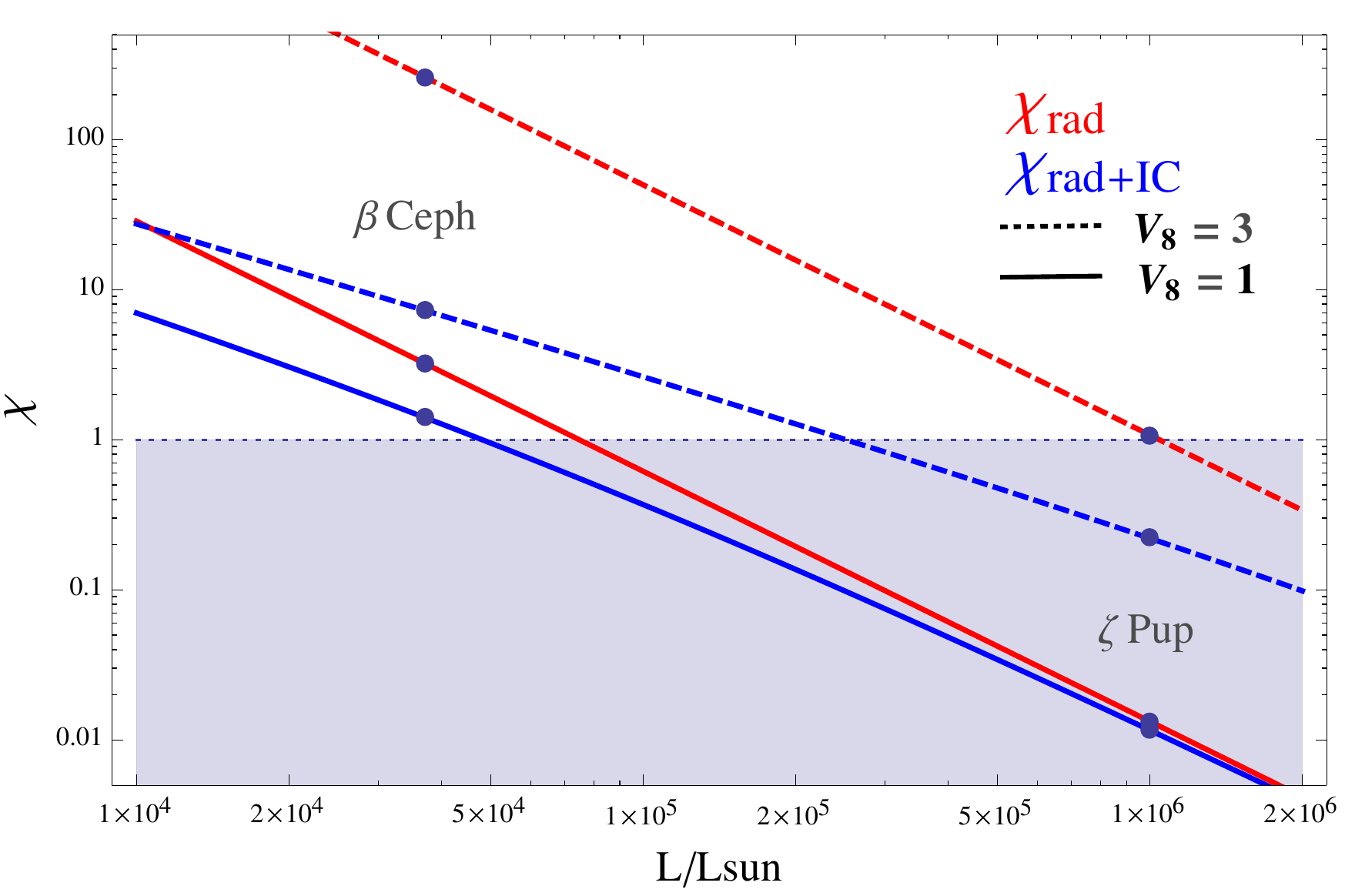}
\caption
{Cooling parameter  for radiative processes ($\chi_{rad}$; red curves) and combined radiative and inverse-Compton processes ($\chi_{rad+IC}$; blue curves), plotted vs.\ stellar luminosity (in solar units $L/L_\odot$), which serves as proxy for increased mass loss rate (${\dot M} \sim L^{1.7}$).
Results are plotted for pre-shock wind speeds $V_8=1$ (solid line) and $V_8=3$ (dashed lines), with fixed radius $r_{12}=1$.
The points give values appropriate to an early B-type star like $\beta$\,Cephei (left) and an O-type supergiant like $\zeta$\,Pup (right).
}
\label{fig:tscales}
\end{center}
\end{figure}

To compare the radiative and IC cooling times, let us relate the mass loss rate and luminosity through the CAK mass loss scaling. Ignoring for simplicity the dependencies on other stellar parameters like stellar mass or radius, we can approximate this as \citep[see, e.g., \S 4.3 of ][]{Owocki04b}
\beq
\Mdot_{-6} \approx  3.6 L_6^{1/\alpha} \approx 3.6 L_6^{5/3} 
\, ,
\label{eq:mdcak}
\eeq
where the latter relation use a typical CAK power index $\alpha \approx 0.6$.

To examine the effect of including such IC cooling, let us now define an associated IC to adiabatic cooling time $\chi_{\rm IC} \equiv t_{\rm IC}/t_{ad}$. 
The total radiative\,+\,IC cooling parameter is then obtained by inverse sum of the components, 
\beq
\frac{1}{\chi_{rad+IC} } = \frac{1}{\chi_{rad}} + \frac{1}{\chi_{\rm IC}} 
\, .
\label{eq:chiradic}
\eeq
As an example,  figure \ref{fig:tscales} plots the luminosity variation of the radiative cooling and radiative\,+\,IC cooling parameters for pre-shock wind speeds $V_8 = 1$ and $V_8 = 3 $.
High luminosity stars  -- e.g.\ the case of the O-supergiant $\zeta$\,Puppis marked by the right-side dots -- are well into the radiative regime (marked by the shading) $\chi < 1$ for both wind speeds. Moreover, the similar values of the radiative and radiative\,+\,IC curves indicate that, for such high luminosity stars, including IC processes has only a marginal additional effect on the cooling.

In contrast, for lower luminosity stars -- e.g.\ the case of the B2-giant $\beta$\,Cephei -- cooling is in the adiabatic regime, especially for the higher-speed case  $v_8=3$; but the IC cooling now significantly reduces the cooling time compared to the models with just radiative cooling, which are even further in the adiabatic regime.

For this simple model of a standing shock in spherical outflow, the overall conclusion from this timescale comparison is that even strong shocks in luminous stars with large mass-loss rates should be radiatively cooled. 
In less luminous stars with weaker winds, IC can significantly enhance cooling over the purely radiative case, but in the  lowest luminosity stars even their combined effect is less than adiabatic expansion.

For X-ray emission from MCWS that form in closed magnetic loops, it is helpful to translate these timescales to associated cooling lengths.
For any post-shock cooling timescale $t$, the associated length scale can be approximated by its product with the post-shock flow speed, $\ell = t V_w/4$.
The ratio of this cooling length to the shock radius is thus $\ell/r_s = t V_w/4 r_s = (3/4) t/t_{ad} = 0.75 \, \chi_{rad}$.
For cases with efficient cooling, $\chi_{rad} \ll 1$, the shock radius should be a small cooling length $\ell$ below the loop apex near the  Alfv\'{e}n radius, implying $R_A \approx  r_s + \ell \approx r_s (1 + 0.75  \chi_{rad})$.

But for inefficient cooling cases with $\chi >1$, the cooling length becomes comparable to the loop apex radius, forcing the shock retreat and associated shock weakening.
As basis for interpreting such shock retreat effects in the MHD simulations below (\S\S 3-4), let us next illustrate this process through an analytic scaling  for this simple  example of a spherical
standing shock.
Appendix B generalizes this to account for the curved flow geometry of material trapped in closed dipole loop.

\subsection {Spherical Scaling for Cooling-Regulated  Shock-Retreat}
\label{sec:shockretreat}

The above scaling analysis  characterizes the efficiency of post-shock cooling by comparing the  timescales in the immediate post-shock transition, focusing particularly on the relative values of the radiative and IC cooling to the expansion timescale assuming a {\em constant} post-shock speed $v= V_w/4$. 
More realistically, for one-dimensional flow against some fixed barrier or ``wall'',  this post-shock speed must slow to zero at this wall,  which in
this simplified spherical expansion model acts as a proxy for the apex radius $r_m$ of a given closed magnetic loop.

The issue at hand then is to derive scalings for the total length $r_m - r_s$ for the cooling layer between this apex and the shock at radius $r_s$. Moreover, to be self-consistent, this should take into account  the radial scaling for the pre-shock wind speed.
As an alternative to solving the full dynamical acceleration of outflows along such a closed magnetic loop, let us simply assume that at any given radius $r$, the flow speed $v$ can be approximated by a standard ``beta'' velocity law,
\beq
v(r) = \vinf ( 1 - \Rstar/r )^\beta \equiv \vinf w(r)
\, ,
\label{eq:blaw}
\eeq
where $w$ represents a scaled speed in terms of the terminal  speed $\vinf$, which for simplicity we take here to have a value equal to that for the non-magnetized wind.
Any flow extending to the apex radius $r_m$  reaches a scaled speed $ w_m \equiv w(r_m) $;
but in general the limited cooling implies a shock retreat to some radius $r_s \le r_m$, with a reduced scaled pre-shock speed $w_s \equiv w(r_s) \le w_m$.
As the cooling becomes more inefficient, the larger cooling layer forces a shock retreat to a lower shock radius with a lower wind speed, for which the shocks are weaker and so have a smaller cooling length. 

\begin{figure}
\begin{center}
\vfill
\includegraphics[scale=0.65]{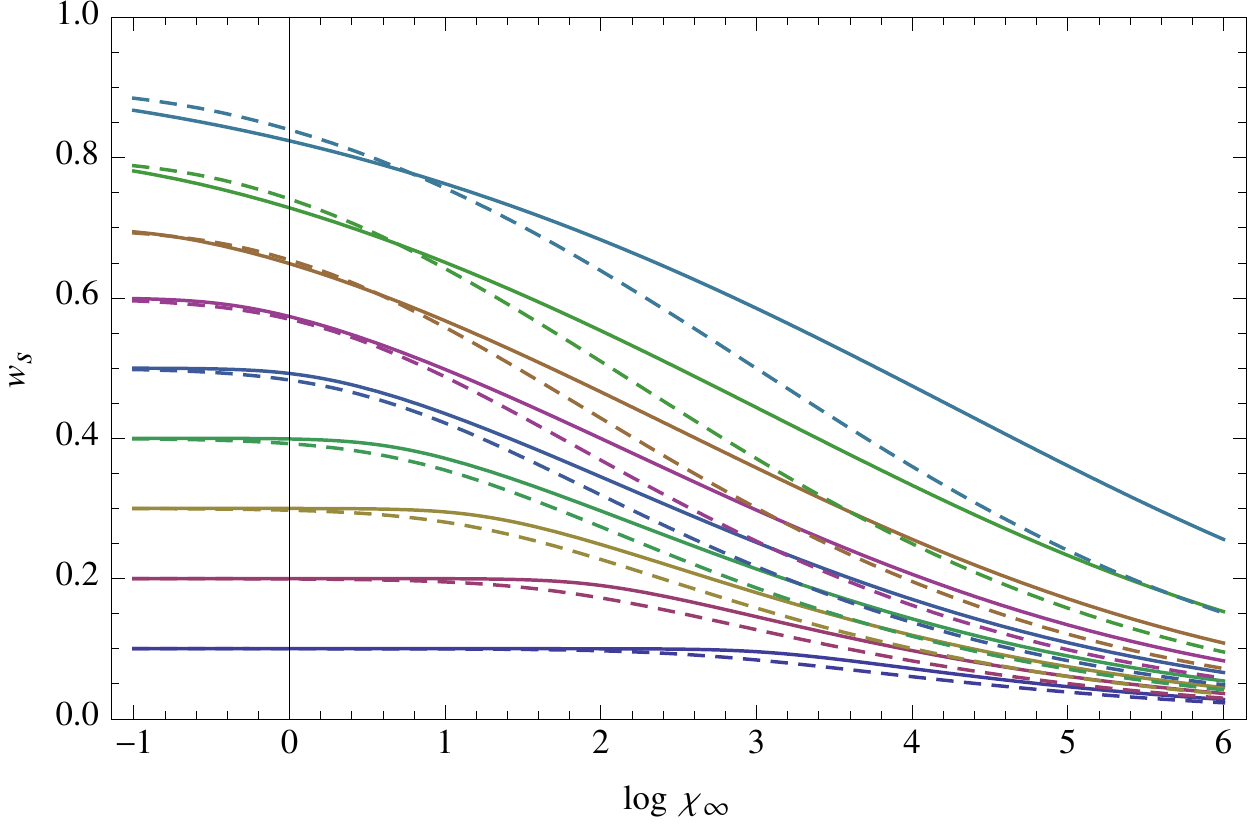}
\caption
{
Reduced shock speed $w_s$ vs. log of the cooling parameter $ \chi_\infty$, plotted for various values of scaled apex speed $w_m $ from 0.1 to 0.9.
Dashed curves are for the simple spherical expansion form for shock retreat, while the solid curves account for dipole loop geometry, as described by the generalized shock-retreat analysis in Appendix \ref{sec:nonsph}.
}
\label{fig:wsvschi}
\end{center}
\end{figure}

To  derive the shock strength that results from this self-regulation by cooling-efficiency shock retreat, let us first solve for the evolution of the post-shock temperature using the steady-state form for the temperature equation (\ref{eq:dTdt}), 
\beq
\frac{v}{T} \frac{dT}{dr} \approx 
%\frac{2}{3} \nabla \cdot {\bf v} +  
\frac{2}{3} \frac{v}{\rho} \, \frac{d \rho}{dr}
%\frac{2}{3} \frac{\mubar}{\mu_e \mu_p} \rho \frac{\Lambda(T)}{kT} +
- \frac{2 \mubar }{3 k} \, \frac{ \rho \Lambda_m }{T} 
\, ,
%\frac{8}{3}  \frac{\kappa_e \mubar }{m_e c}  \,  U_{ph }
\eeq
where for simplicity we have neglected  IC cooling.
Here we have used the steady-state mass continuity to rewrite the adiabatic cooling in terms of the velocity and density.
Since the post-shock flow is by definition subsonic, we can approximate it as nearly {\em isobaric}, implying that $\rho T \approx \rho_s T_s$, where the post-shock temperature is given by eqn.\ (\ref{eq:tsdef}) and the  post-shock density by $\rho_s = 4 \rho_w = \Mdot/(\pi V_w (r_s) r_s^2)$.
Using this and the mass continuity  to eliminate both the speed $v$ and density $\rho$ in favor of the temperature $T$, we can combine the adiabatic cooling with the advection along the temperature gradient, leading to a simple first-order differential equation for the post-shock temperature,
\beq
\frac{T^2}{r^2} \, \frac{dT}{dr} =  - \frac{4/5}{\chi_{rad}} \,  
%\left ( \frac{T_s}{r_s} \right )^3
 \frac{T_s^3}{r_s^3} 
\, ,
\label{eq:dtdrshock}
\eeq
where the  factor $4/5$ adjusts for constants used in the above definition (\ref{eq:chirad}) for the cooling parameter $\chi_{rad}$ associated with the cooling time from an assumed adiabatic expansion.
With the boundary condition $T(r_s) = T_s$, 
eqn.\ (\ref{eq:dtdrshock})  can be trivially integrated to give an explicit solution for temperature in the post-shock region\footnote{For simplicity, this assumes a constant cooling parameter $\Lambda_m$. It is trivial to extend the analysis to a power-law temperature variation. For example, the rough fit $\Lambda_m \sim T^{-1/2}$ gives a scaling in which the exponent value $1/3$ in the derived solution  (\ref{eq:trpshock}) is replaced by $2/7=1/3.5$.},
\beq
% T(r) = T_s \, \left [ K_s  \left ( 1 - \left ( \frac{r}{r_s} \right )^3 \right ) + 1 \right ]^{1/3}
 T(r) = T_s \, \left [ \frac{4/5}{\chi_{rad}}  \left ( 1 - \left ( \frac{r}{r_s} \right )^3 \right ) + 1 \right ]^{1/3}
\,.
\label{eq:trpshock}
\eeq
Identifying the loop apex radius $r_m$ as  a barrier location where the temperature formally drops to zero, $T(r_m) \equiv 0$, we find\footnote{Note  that in the strong cooling limit $\chi_{rad}  \ll 1$,  this gives $r_m  \approx r_s + \ell = r_s (1 + 5 \chi_{rad}/24)$, implying a  cooling length $\ell $  that is a factor $5/18 \approx 0.28$ smaller than the value $ 0.75 r_s \chi_{rad}$ predicted at the end of \S \ref{sec:sphshock}.
This correction reflects the significant deceleration of the post-shock flow speed, with associated increases in density, both of which lead to stronger cooling and so a shorter cooling length than predicted by a simple constant-speed advection over the  post-shock timescale.}

\beq
%\frac{r_m}{r_s} = \left ( 1 + \frac{\chi_{rad}}{k_s} \right )^{1/3}
\frac{r_m}{r_s} = \left ( 1 + \frac{5 \chi_{rad}}{4} \right )^{1/3}
\, .
\label{eq:rwbrs}
\eeq
We can readily turn this around to solve for the shock radius, accounting for the fact that, from (\ref{eq:chirad}),  $\chi_{rad} \sim r_s V_w^4 (r_s) $. 
Assuming a simple $\beta=1$ velocity law (\ref{eq:blaw}), we find
%for $r_s$,
\beq
\left ( \frac{r_m}{r_s} \right )^3  = 1 + { \chi_\infty}\, \frac{r_s}{\Rstar} \left ( 1- \frac{\Rstar}{r_s} \right )^4 
\, ,
\label{eq:implicitrs}
\eeq
which alternatively can be cast as an equation  
for the scaled shock speed $w_s$,
\beq
\left ( \frac{1-w_s}{1-w_m} \right )^3  = 1 +  \chi_\infty \, \frac{w_s^4}{1-w_s} 
\, .
\label{eq:implicitws}
\eeq
Here we have defined a cooling parameter associated with  the terminal speed, $v=V_\infty$, evaluated at the 
stellar radius $\Rstar$, while also absorbing the $5/4$ factor,
\beq
\chi_\infty  \equiv  \frac{15 \pi  }{128} \, \frac{V_\infty^4 \Rstar} { \Mdot  \Lambda_m  }
\approx
0.034  \, \frac{V_{8}^4 \, R_{12} }{\Mdot_{-6}}
\, .
\label{eq:chiinf}
\eeq
The numerical evaluation uses the scaled values $V_{8} \equiv V_\infty/(10^8$cm/s) and $R_{12} \equiv  \Rstar/10^{12}$\,cm.  For typical values $V_8 = 3$ and $R_{12} = 1$, $\chi_\infty = 1$ corresponds to a wind mass loss rate $\Mdot_{-6} \approx $\,0.8.
Comparison with eqn.\ (\ref{eq:chirad}) for $\chi_{rad}$ shows a superficially similar scaling to (\ref{eq:chiinf}) for $\chi_\infty$; but it is important to note that $\chi_{rad}$ represents a comparison between radiative to adiabatic timescales at some {\em local} shock radius $r_s$, while $\chi_\infty$ is a fixed {\em global} characteristic of the star that controls the spatial shock retreat.

Given $\chi_\infty$, and the apex speed $w_m$,
eqn.\ (\ref{eq:implicitws}) can be readily solved for $w_s$ by standard root finding. 
For this simple spherical example of shock retreat, the dashed curves in
figure \ref{fig:wsvschi} plot $w_s$ vs. $\log \chi_\infty$ for a range of $w_m$.
The solid curves compare results for the generalization derived in Appendix \ref{sec:nonsph}
to account  (though solution of eqn.\ (\ref{eq:wsdip})) for the dipole loop geometry.

Associating the maximum loop radius with the Alfv\'{e}n radius, which scales with the magnetic confinement as $R_A \sim \eta_\ast^{1/4}$, we can use this dipole shock retreat solution to estimate the reduction in shock temperature $T_s $,  and thus the reduced shock energy dissipation available for X-ray emission. \S \ref{sec:xadm} develops this further to derive analytic scaling laws for $L_x$ as function of $\eta_\ast$ and $\chi_\infty$.
This proves very helpful for interpreting results from the full numerical MHD models that we now describe.

\begin{figure*}
\begin{center}
\vfill
\includegraphics[scale=0.148]{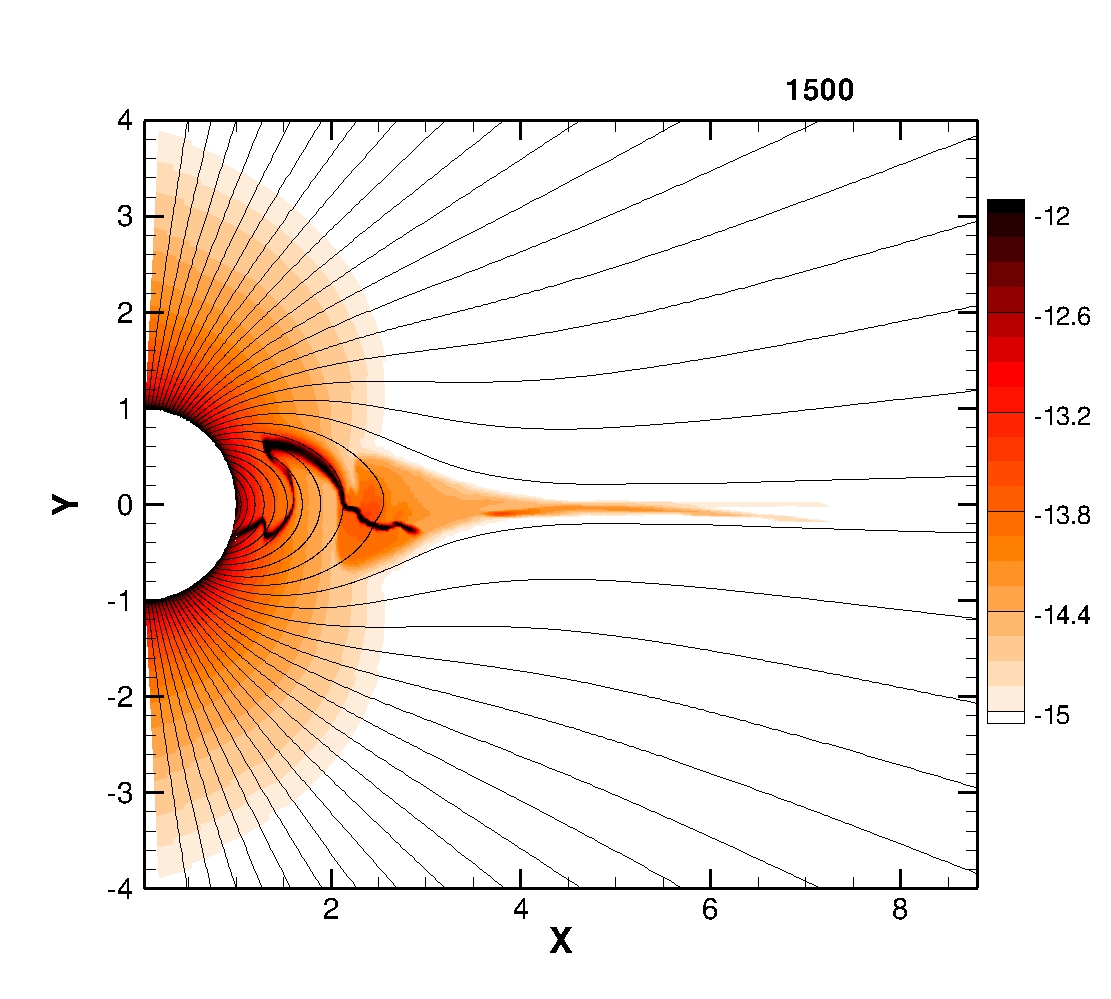}
\includegraphics[scale=0.148]{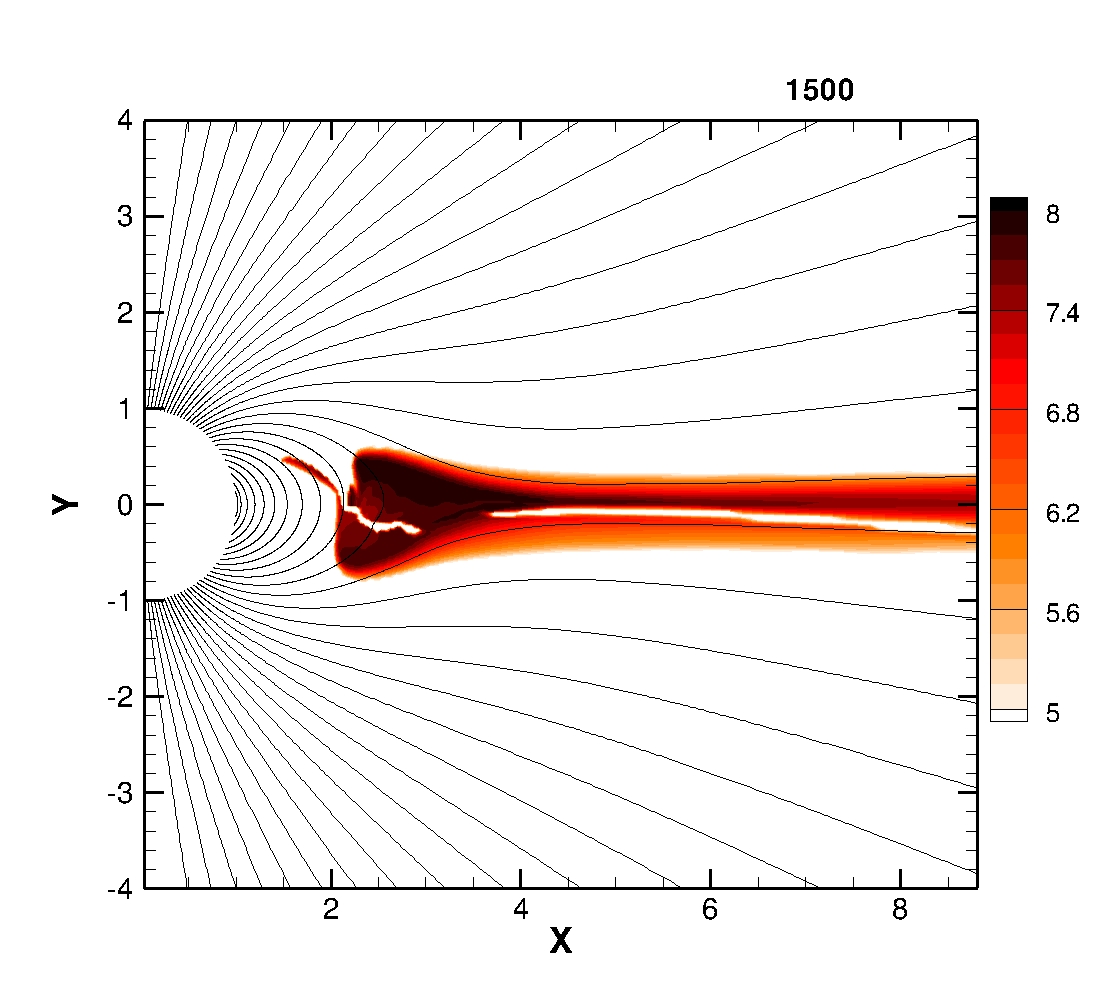}
\includegraphics[scale=0.148]{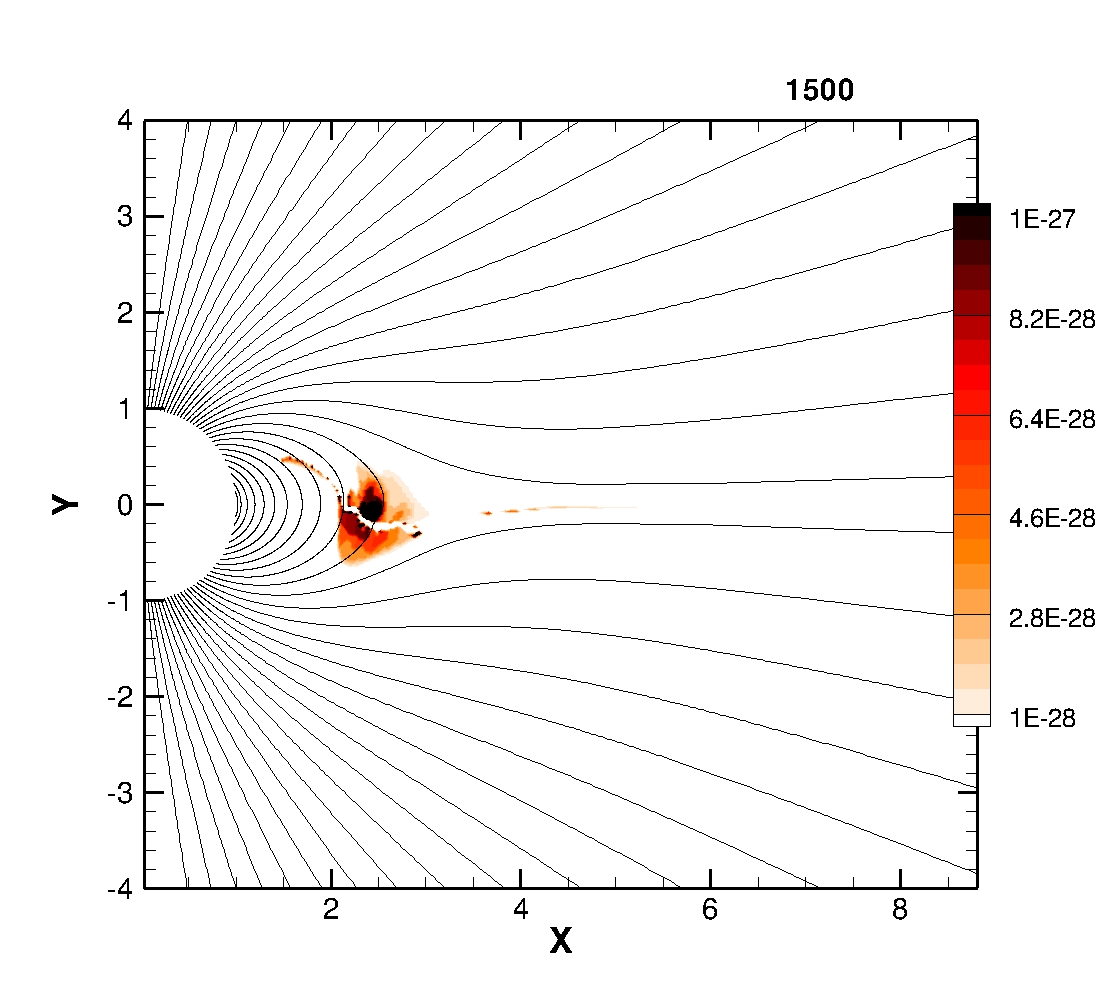}
\caption{
Color plots of log density (left) and log temperature (middle) for arbitrary snapshot of structure in the standard model with $\eta_\ast =100$ and no IC cooling.
The right panel plots the proxy X-ray emission $XEM_{T_x}$  (weighted by the radius $r$) from (\ref{eq:xemdef}),  on a {\em linear} scale for a threshold X-ray temperature $T_x = 1.5$\,MK.
}
\label{fig:rhotxem}
\end{center}
\end{figure*}

\section{MCWS X-RAYS FROM STANDARD MODEL}
\label{sec:stdmod}

\subsection{Model description and parameters}
\label{sec:stdmodparams}

Let us now turn to our numerical simulations of shock heating and X-ray emission in MCWS.
As a basis for our study of how cooling efficiency affects X-ray emission, let us first examine the X-ray properties for the same standard model that formed the basis of the previous MHD parameter studies in papers I-III.

Roughly representative of an O-type supergiant star like $\zeta$\,Puppis, this model assumes
a radius $R_{\ast} = 19 \Rsun$, luminosity $L = 10^6 \Lsun$,
and an effective mass of $M= 25\Msun$.
(This reflects a factor two reduction below the Newtonian mass to  account
for the outward force from the electron scattering continuum.)
Within the standard, finite-disk-corrected CAK model, in a non-magnetic star this leads to a mass loss rate  $\Mdot \sim 3.3 \times 10^{-6} \Msun$/yr and wind terminal speed $V_\infty \approx 3000\,$km/s.
As illustrated in fig. \ref{fig:tscales}, this model is generally within the cooling regime $\chi < 1$, with IC making only a minor contribution to the overall cooling, except for high wind speeds $V_8  \sim 3$.

Our standard model assumes a magnetic  confinement parameter  $\eta_\ast =100$,  giving then an Alfv\'{e}n radius $R_{\rm A}/\Rstar \approx \sqrt{10} \approx  3.1$. For the  stellar and wind parameters quoted above, this requires a polar magnetic field of $B_p = 3$\,kG.   
Since these stellar and wind  parameters are fixed throughout this paper, exploration of any models with different $\eta_\ast$ is done simply by adjusting the assumed dipolar field strength by the prescription, $B_p = 300 {\rm G} \sqrt{\eta_\ast}$.
Specifically, the models below with $\eta_\ast = 10$ assume $B_p \approx 1000$\,G.

For all  simulations here, the numerical specifications -- such as the computational grid, initial condition, and
boundary conditions -- are  as in Paper I. The initial condition  introduces the dipole field of chosen strength into a relaxed steady, spherically symmetric wind driven by line-scattering of stellar radiation according to the CAK formalism.  The temperature is initially set to the stellar effective temperature $T_{\rm eff}$, but now varying according to the energy equation (\ref{eq:energy}) to allow for shock-heating and post-shock cooling, keeping however a floor at $T_{\rm eff}$ as a proxy for the photoionization heating by the stellar UV radiation.
To average over dynamic structure associated with wind trapping and infall, the models are run to a maximum time $t_{\rm fin} $ that is many times the wind flow time $t_{\rm flow} = R_{max} V_\infty \approx 150$\,ks over the model range extending to $R_{max} = 15 \Rstar$.
For the standard model, we take $t_{\rm fin} =$\,3000\,ks, but for the broader parameter study we use a common value that is half this standard, i.e., $t_{\rm fin} =$\,1500\,ks.
To allow for relaxation from the initial condition, all quoted time-averaged quantities here are computed starting at $t=500$\,ks, and extending to $t_{\rm fin} $\,ks.

\begin{figure*}
\begin{center}
\vfill
\includegraphics[scale=0.20]{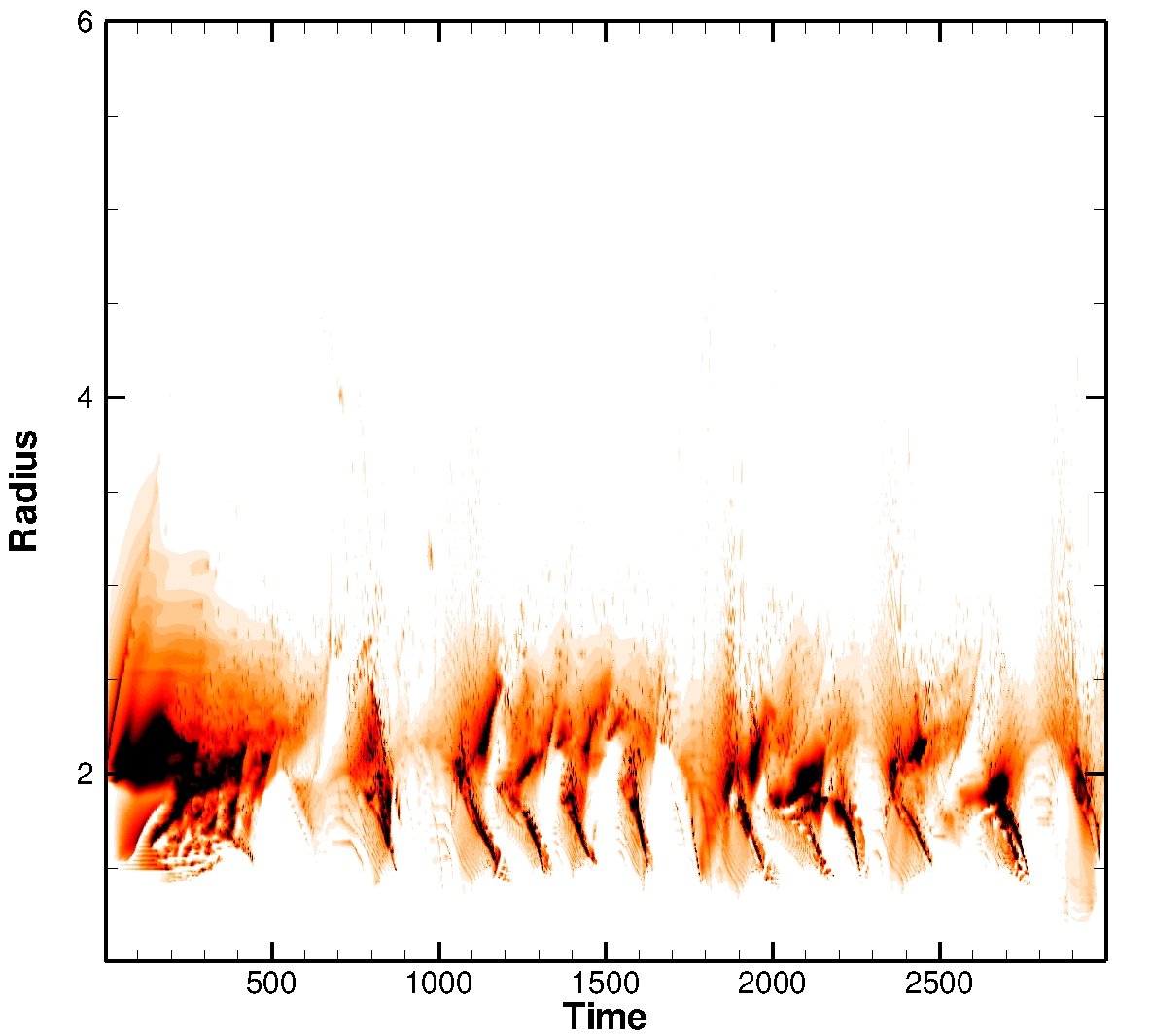}
\includegraphics[scale=0.20]{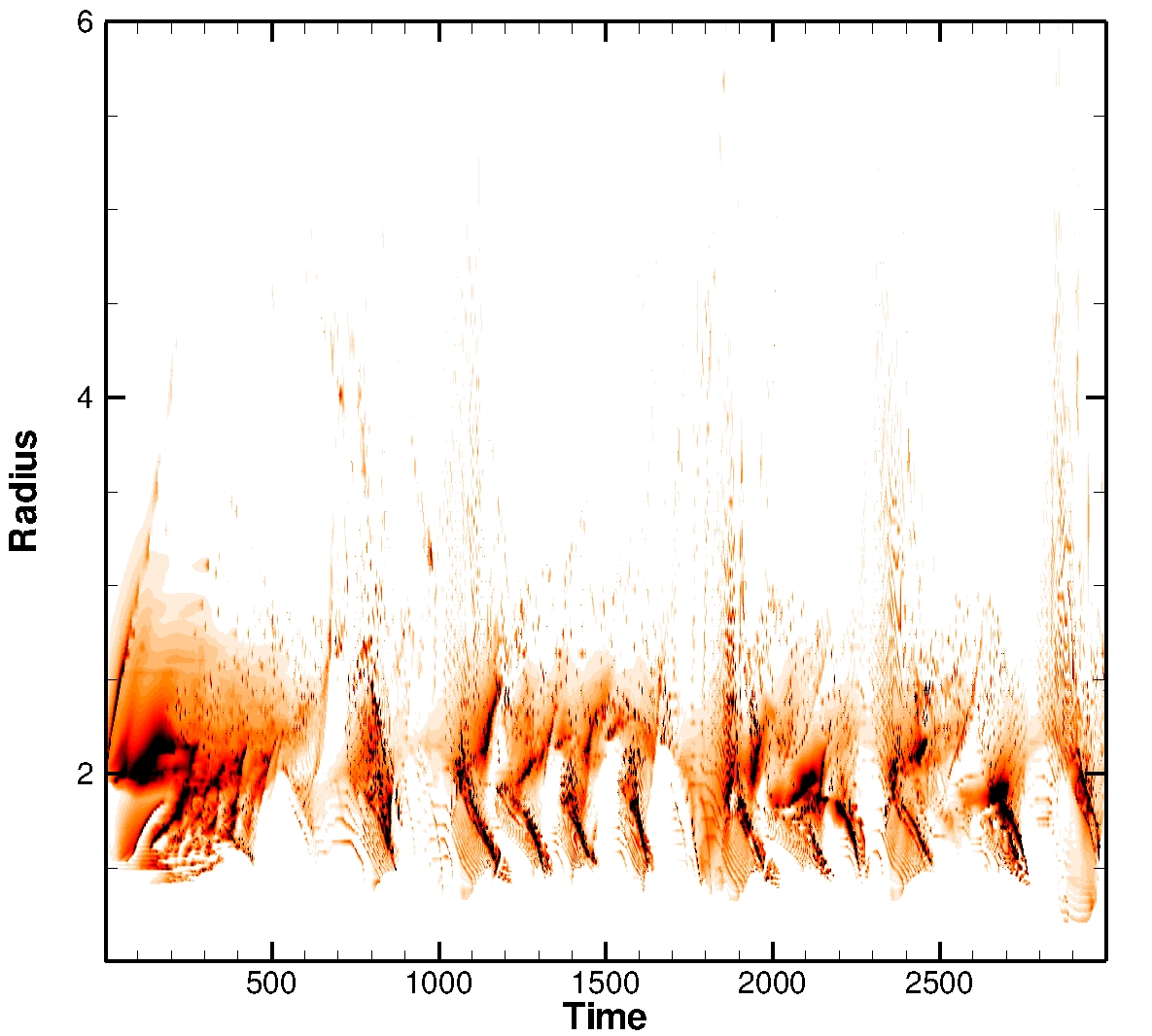}
\caption{
Radial distribution of latitudinally integrated X-ray emission above an X-ray threshold, plotted vs.\ radius (in $\Rstar$) and time (in ks) for the standard model.
The left panel uses the Boltzmann formula (\ref{eq:xemdef}) with $T_x=$1.5\,MK, and right panel shows the actual energy-integrated X-ray emission above a threshold $E_x =$\,0.3\,keV.
}
\label{fig:dxemdr}
\end{center}
\end{figure*}

\subsection{Density and Temperature structure and associated X-ray Emission}

Let us first consider a model with radiative cooling, but ignoring IC cooling, and with a moderately strong magnetic confinement $\eta_\ast = 100$, implying an Alfv\'{e}n radius $R_A \approx 3.1 \Rstar$.
The left and middle panels of figure \ref{fig:rhotxem} show color plots of  the  characteristic spatial structure in log density and log temperature at a fixed time snapshot, chosen arbitrarily here to be half the final time $t=t_{\rm fin}/2=$1.5\,Ms.
Note that the highest density occurs in radiatively cooled regions with low temperature (near the floor at $T \approx T_{\rm eff}$), while the shock-heated regions with temperatures up to $\log T \approx 7.5$ (K) have relatively low density.

To characterize the regions of X-ray emission, which scales with the density-squared emission measure of material that is hot enough to emit X-rays, let us define a simple proxy that weights the emission measure by a Boltzmann factor for some threshold temperature $T_x$,
\beq
X_{T_x}  (\rho, T) \equiv  \rho^2 \exp (-T_x/T)
\, .
\label{eq:xemdef}
\eeq
The rightmost panel of figure \ref{fig:rhotxem} shows a color scale plot of
 $X_{T_x}$ for a threshold temperature $T_x=1.5$\,MK, sufficient to produce X-rays of $\sim$0.1\,keV and above.
Note that the X-ray emission is concentrated near the top of the outermost closed loop, just below the Alfv\'{e}n radius, $R_A \approx 3.1 \Rstar$.
This is much more localized than the distributed regions of high temperature, which extend outward well beyond the Alfv\'{e}n radius, centered on the current sheet that defines the jump in polarity for wind-opened field line on each side of the magnetic equator.
While impressive in a color plot of the temperature, such extended regions have too low a density to produce much significant X-ray emission.

\subsection{Radius-time plots of latitudinally integrated X-ray emission}

Such snapshots do not capture the extensive  dynamical variability that is inherent from the trapping and subsequent infall of material in closed magnetic loops,  as can be seen by animations of the evolving structure.

To capture this here in a still graphic, let us collapse one of the spatial dimensions by latitudinally integrating this X-ray emission measure XEM\footnote{This is analogous to the latitudinally integrated mass distribution defined to illustrate the $r,t$ accumulation of equatorial mass in the rotating wind models of paper II. See figs.\ 4, 5, 7 and 9 there.},
\beq
\frac{d\, {\bar X}_{T_x}}{dr} (r,t) \equiv 2\pi r^2  \int_{-\pi/2}^{\pi/2}  \sin (\theta) X_{T_x}  [ \rho (r,\theta,t),T (r,\theta,t) ] \, d\theta
\, ,
\label{eq:dxemdrdef}
\eeq
For this standard model with $\eta_\ast =$100  (and neglecting IC cooling), the left panel of figure \ref{fig:dxemdr} then shows color plots of the time and radius variation of this integrated XEM for the threshold temperature, $T_x$=\,1.5MK.

The right panel shows the actual distribution of total X-ray emission above an energy threshold $E_x =$0.3\,keV, computed using the spectral synthesis method described in the next section.
The close correspondence supports the utility of the simple Boltzmann form (\ref{eq:xemdef}) for characterizing the total X-ray emission.

But both plots provide a vivid illustration of the intrinsic time variability and spatial structure of the X-ray emission in such MHD simulation models.  
Quickly after the start-up condition, strong initial shocks form to produce extensive 
X-ray emission, centered on a radius $r \approx 2.1 \Rstar$, but extending from $r=1.5 \Rstar$ up to around $r \approx 2.8 \Rstar$, i.e.\ just below the Alfv\'{e}n radius $R_A \approx 3.1 \Rstar$. By $t=500$\,ks the cooling and infall of this shock-heated material leads to a brief interval of weak emission, which however recovers as new, somewhat less organized and thus somewhat less distributed shock heating with more moderate X-ray emission. This material again cools and leads to repeated cycles of shock-heated X-ray emission and low-emission infall, with quasi-regular period about $250$\,ks.

While quite distinctive in the 2D simulations here, in more realistic 3D models the likely phase incoherence among heating/infall cycles at different azimuths would tend to smooth out any overall variability in observed X-rays. In the 3D model computed in \citet{Uddoula13}, for example, such azimuthal averaging greatly reduces the stochastic variations derived for Balmer line emission.

\begin{figure*}
\begin{center}
\vfill
\includegraphics[scale=0.65]{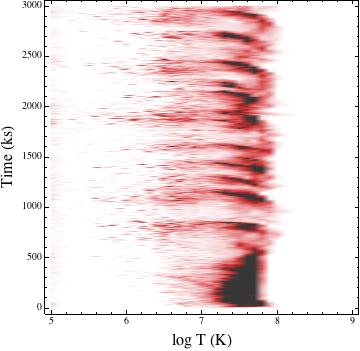}
\includegraphics[scale=0.65]{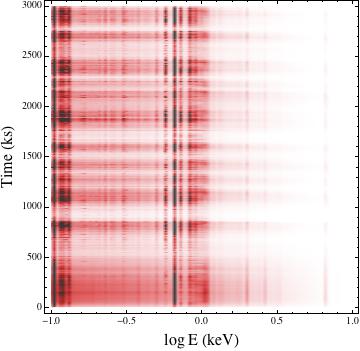}
\caption{
{\em Left:}
Differential emission measure, $ DEM (t,T)$, plotted with a linear color scale versus time $t$ (in ks) and log temperature $\log T$ (in K) for the standard model without IC cooling.
{\em Right:} Associated dynamic X-ray spectrum $ L_x (E,t)$, plotted with a linear color scale vs.\ time and $\log E$ (in keV).
}
\label{fig:demstdmod}
\end{center}
\end{figure*}

\begin{figure*}
\begin{center}
\vfill
\includegraphics[scale=0.67, angle=0]{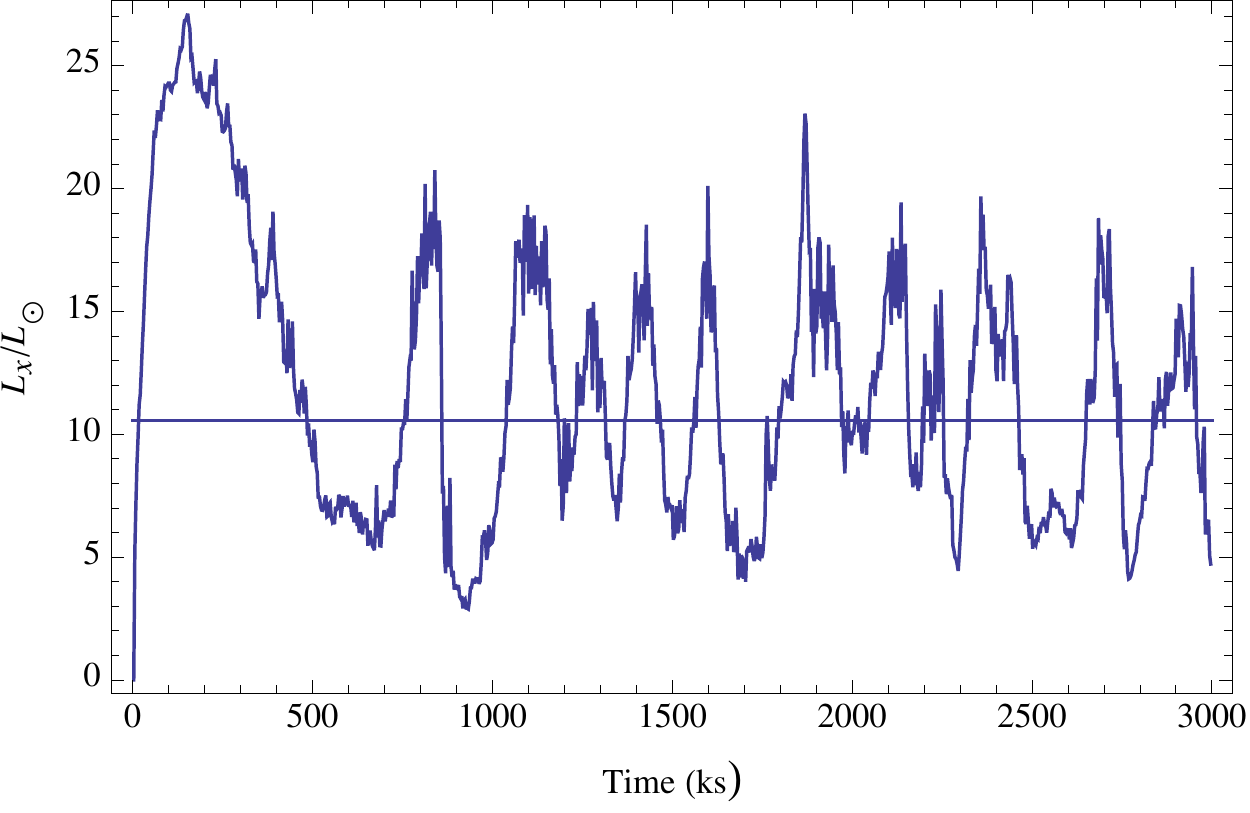}
\includegraphics[scale=0.67]{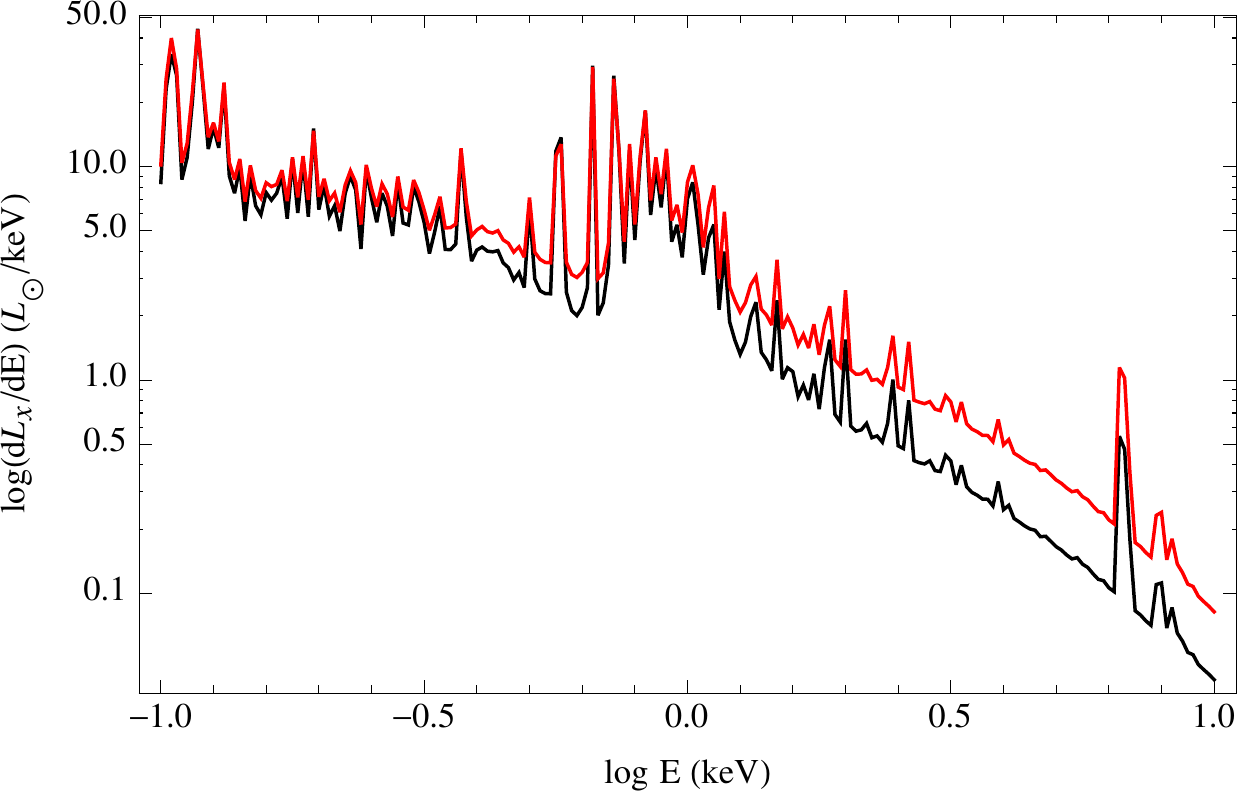}

\caption{
{\em Left:} For standard model simulations with $\eta_\ast=100$, the time variation of cumulative X-ray luminosity $L_x (E>E_x,t) $ above X-ray threshold energy $E_x $=\,0.3\,keV, plotted in units of $\Lsun$.
 The horizontal line shows the time-averaged value $\ L_x  \approx 67 \Lsun$, computed over times $t>500$\,ks, after the model has relaxed from its initial condition.
{\em Right:}
Log scale of time-averaged X-ray spectrum $dL_x /dE$ vs.\ $\log E$.
The black and red curves compare results with and without IC cooling.
}
\label{fig:lxelxt}
\end{center}
\end{figure*}

\subsection{Dynamic spectrum}

Let us now examine the dynamic X-ray spectrum that arises from this cycle of shock-heating and mass infall.

The X-ray emission at any photon energy $E$ can be computed using the energy-dependent emission function, $\Lambda_m (E,T)$, derived from a standard plasma emission code  like the APEC model \citep{Smith01, Foster12} in XSPEC
\citep{Arnaud96}.

Integration over all energies gives the total cooling function introduced in  eqn.\ (\ref{eq:qrad}), $\Lambda_m (T) = \int \Lambda_m (E, T) \, dE$.
The energy-dependent volume emissivity (with CGS units erg/(cm$^3$\,s\,keV)) just weights this by the associated density-squared emission measure (EM) of gas at the given temperature,
\beq
\eta_x (E,\rho, T) = \rho^2 \Lambda_m (E,T) 
\, .
\label{eq:etaerhot}
\eeq
Integration over the full spherical volume of the model then gives (neglecting any absorption or occultation) the energy spectrum of total emitted luminosity,
\beq
L_x (E) 
=  \int  \Lambda_m (E,T) \rho^2 \, dV  
\equiv \int  \Lambda_m (E,T) \frac{d \, EM(T)}{d \ln T} d \ln T
\, ,
\label{eq:demdef}
\eeq
where the latter equality defines the volume-integrated {\em differential emission measure}, DEM $\equiv d \, EM(T)/d \ln T$.

The color plots in  figure \ref{fig:demstdmod} illustrate the time variations of the DEM$(t,T)$ (vs.\ $\log T$, left) and the resulting dynamic X-ray spectrum $L_x(E,t)$ (vs.\ $\log E$, right).
Note again the dynamical variability from the trapping and subsequent infall of material in closed magnetic loops.

 \begin{figure*}
\begin{center}
\vfill
\includegraphics[scale=0.67]{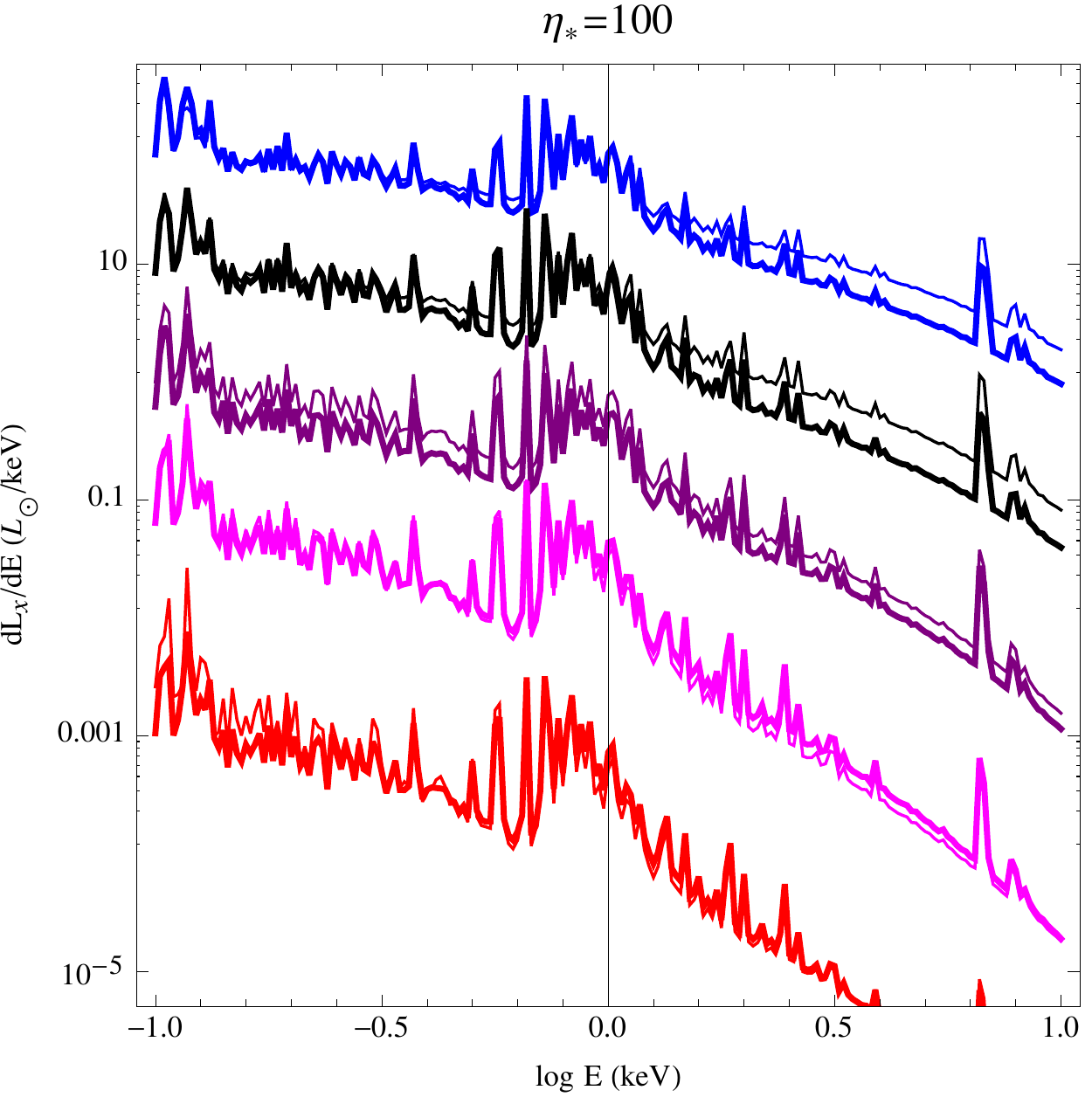}
\includegraphics[scale=0.67]{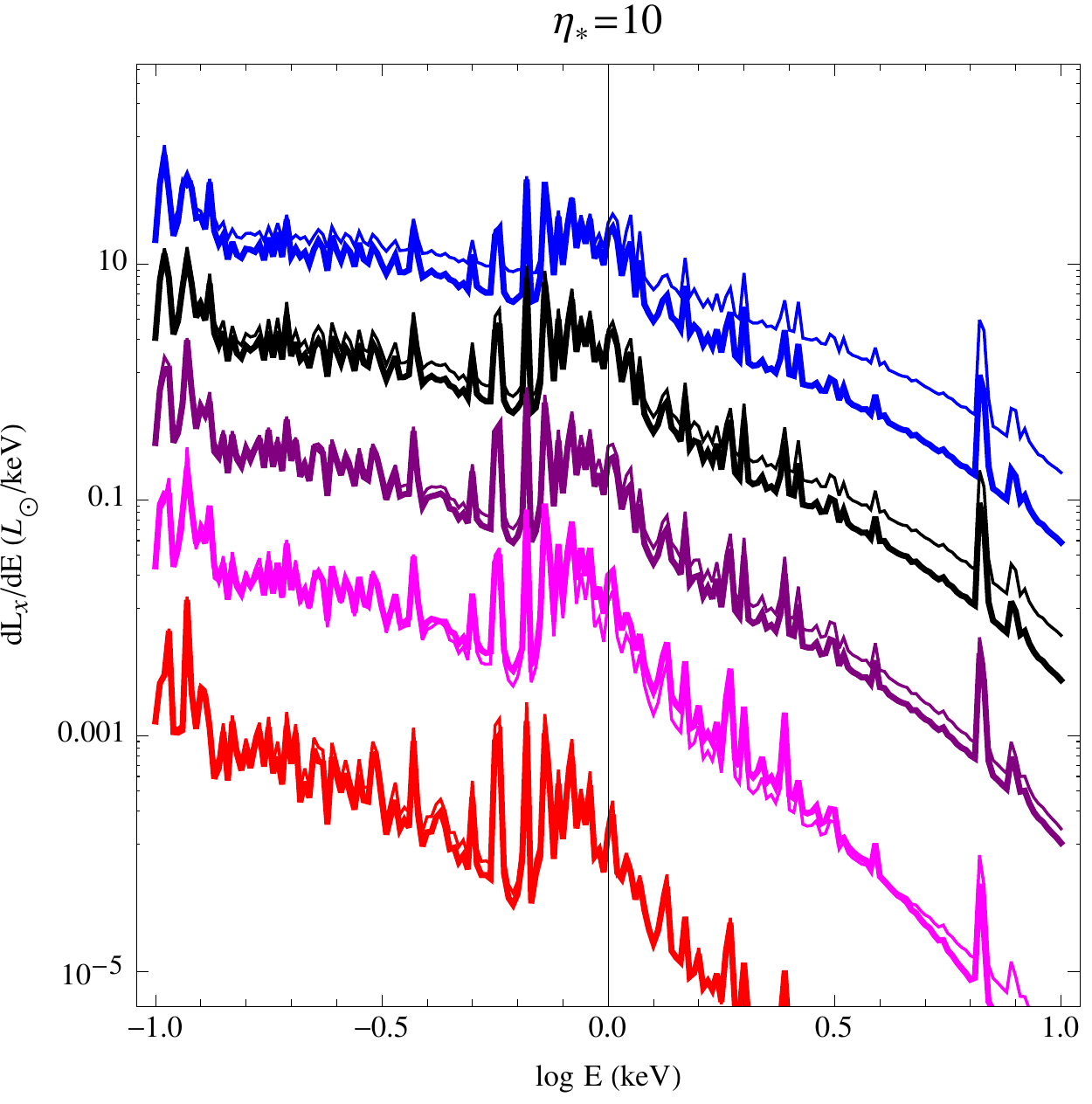}
\caption{
{\em Left:} Time-averaged luminosity spectra $L_x (E)$ vs.\ $\log E$ for $\eta_\ast=100$ models, plotted on a log scale in units of $\Lsun$ for the full series of 5 models with cooling efficiencies $\epsilon_c = 10^{-3}$ (lowermost curves, in red) to $10^{+1}$ (top, in blue) in steps of 1 dex.  The thick line curves include IC cooling, while the thinner curves are for radiative cooling only.
{\em Right:} Same as left panel, but for $\eta_\ast = 10$.
Since  EM $\sim \Mdot^2$, and the $\epsilon_c$ is a proxy for $\Mdot$, the $L_x$ values are scaled here by 
$\epsilon_c^2$ from what is derived from the numerical computation with the fixed parameters of the standard model shown by the  black curves. 
}
\label{fig:lxe-rad-vs-ic}
\end{center}
\end{figure*}

\begin{figure*}
\begin{center}
\vfill
\includegraphics[scale=0.65]{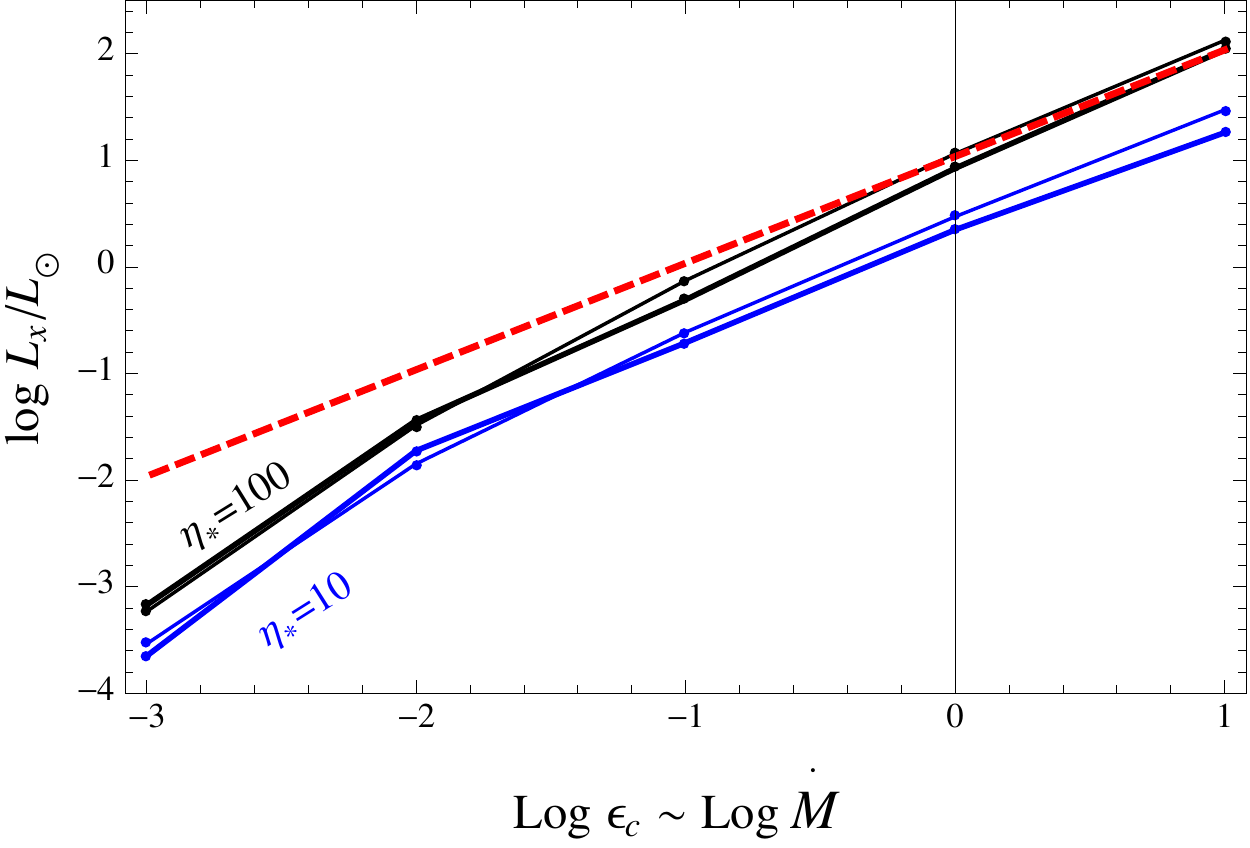}
$~~$
\includegraphics[scale=0.65]{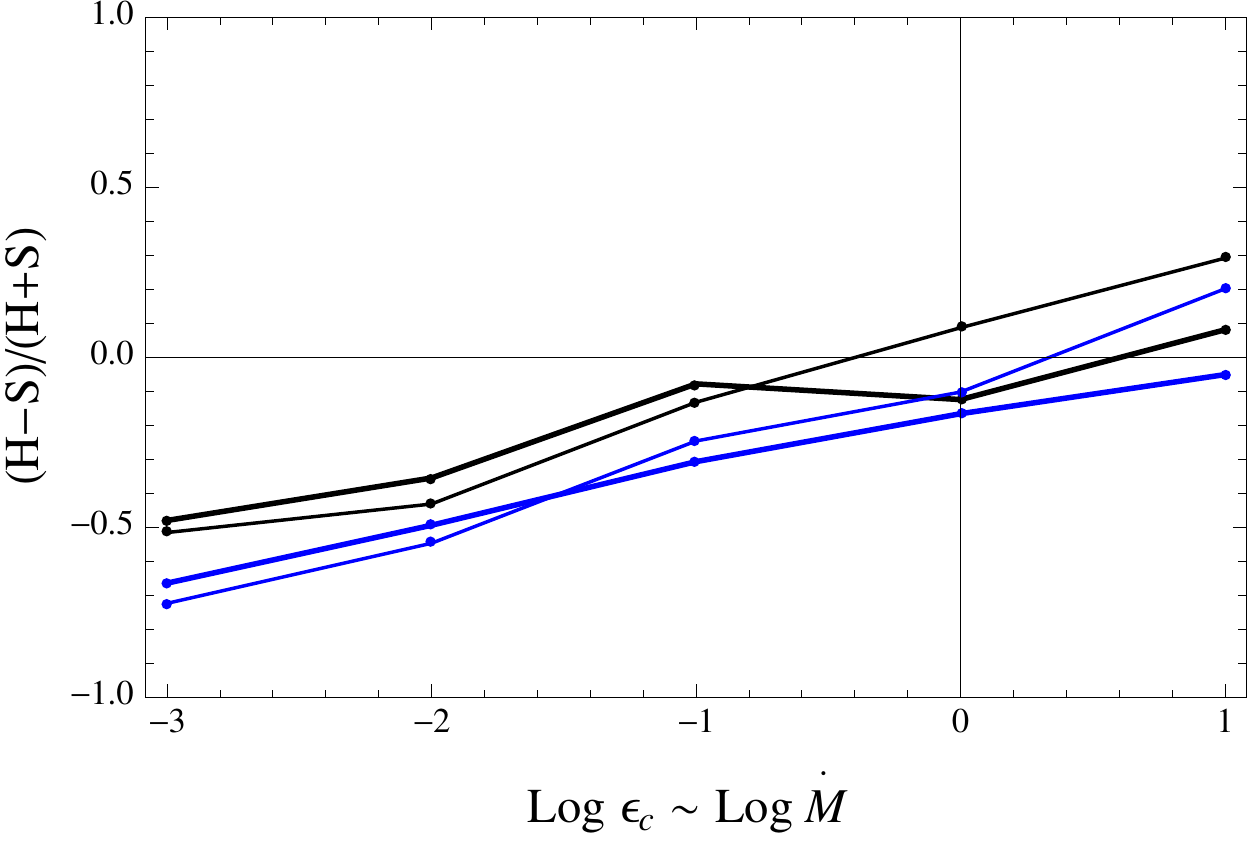}
\caption{
{\em Left:} 
Log of time-averaged X-ray luminosity, $\log L_x$, for X-rays  above $E_x = 0.3$\,keV, plotted vs.\  log of cooling efficiency $\log \epsilon_c $, which acts as a proxy for mass loss rate $\Mdot$. 
The upper  (black) and lower (blue) curves are respectively for $\eta_\ast=$100  and $\eta_\ast =10$, and the thick and normal thickness lines represent models with and without IC cooling.
The dashed red line shows a linear relation normalized to values for the $\eta_\ast =100$ model with the strongest cooling  $\epsilon_c = 10$.
{\em Right:} 
Analogous plots of hardness ratio (H-S)/(H+S) vs. $\log \epsilon_c$, where H represents emission from 1 to 10\,keV, and S represents emission from 0.3 to 1\,keV.
}
\label{fig:lxhrvseps}
\end{center}
\end{figure*}

In these terms, the radius-time variation of total X-ray emission above a threshold, as plotted in the right panel of figure \ref{fig:dxemdr} for $E_x =$0.3\,keV, is defined by
\beq
\frac{dL_{E_x}}{dr} (r,t)
\equiv  \int_{-\pi/2}^{\pi/2} r^2 \sin (\theta)  \rho^2 (r,\theta,t) \Lambar_m [T (r,\theta,t),E_x]  \, d\theta
\label{eq:dlexdr}
\eeq
where 
\beq
\Lambar_m (T,E_x) \equiv \int_{E_x}^\infty \Lambda_m (E,T) \, dE
\, 
\label{eq:lambardef}
\eeq
defines a spectrally integrated emission function. (See Appendix \ref{sec:appendix}.)
The right panel of figure \ref{fig:lamplts} plots $ \Lambar_m (T,E_x) $ vs. $\log T$ for $E_x =$~0.3, 1, and 2~keV, with the dashed lines comparing the corresponding Boltzmann model fits for $T_x = $~1.5, 7 and 20~MK.
The left and right panels of figure \ref{fig:dxemdr} respectively use $T_x=$1.5\,MK and $E_x=$0.3\,keV, giving, as noted, very similar characterizations of the radius and time variation of the associated X-ray emission.

Further integration of (\ref{eq:dlexdr}) over radius give the full volume-integrated X-ray luminosity above the given threshold $L_x  (t) = L_{E_x} (t)$. The left panel of figure \ref{fig:lxelxt} plots this vs.\ time.  The semi-regular episodes of shock-formation and infall lead to a roughly factor 2 variation about the time-averaged value, $\left < L_x \right > \approx 67 \, \Lsun$,  computed over the interval $t=500 - 3000 $\,ks after the initial shock evolution has settled to its quasi-steady state.
The right panel of   figure \ref{fig:lxelxt}  plots   the  time-averaged luminosity spectrum $L_x (E)$ vs.\  $\log E$. The black and red curves compare results with and without IC cooling.
The overall effect is to reduce the hard X-rays, and so soften the spectrum, with however little change in the total emission, which is strongest at lower energies.

\section{Parameter Study for Cooling Efficiency}
\label{sec:cooleff}

\subsection{Varying cooling efficiency as a proxy for variations  in $\Mdot$ and $L$} 

Let us now examine results from an extensive parameter study of MHD simulations with radiative and IC cooling designed to examine how variations in cooling efficiency affect the X-ray emission.

To study the effect on cooling for a lower $\Mdot$ that would be expected from lower luminosity stars, we simply reduce the cooling efficiency in our standard stellar wind model by some fixed factor, $\epsilon_c$, where our study spans a grid of 5 cases with $\epsilon_c$ $= 10^{-3}$ to $10^{+1}$ in steps of 1 dex. In essence, this mimics the effect of changing $\Mdot$ by $\epsilon_c$, while allowing us to keep the magnetic confinement $\eta_\ast$ constant without adjusting the actual field strength. It also avoids the complications of secondary changes in, .e.g., the stellar radius or mass, that would be associated with actual changes in $\Mdot$ in real stars.  (Note that we have included higher $\epsilon_c$ to study the strong cooling limit, even though there are no known magnetic stars with mass loss 10 times the standard $\zeta$ Pup-like case.)  

In models that include IC cooling, we accordingly modify its efficiency by $\epsilon_c^\alpha$, where $\alpha=0.6$ is the CAK exponent.  This is because IC cooling scales with luminosity $L \sim  \Mdot^\alpha$. 
Because this is weaker than the $\Mdot$ scaling of radiative cooling, IC is formally the stronger cooling mechanism for lower-luminosity stars. Moreover, in contrast to radiative cooling, which for higher shock temperatures $T_s$ is reduced by $1/T_s^2$, IC cooling is {\em independent} of $T_s$, and so it tends to be particularly effective in getting cooling started. But as the shock cools, radiative cooling takes over, and so it can never be neglected. 

Overall,  as shown for the above standard case,  IC cooling can reduce the DEM at the highest temperatures; but because its scaling with luminosity generally trends in the same sense as the mass loss scaling of radiative cooling, adding IC has only a modest overall effect on the DEMs and X-ray spectra compared to corresponding models with only radiative cooling.

\begin{figure*}
\begin{center}
\vfill
\includegraphics[scale=0.64]{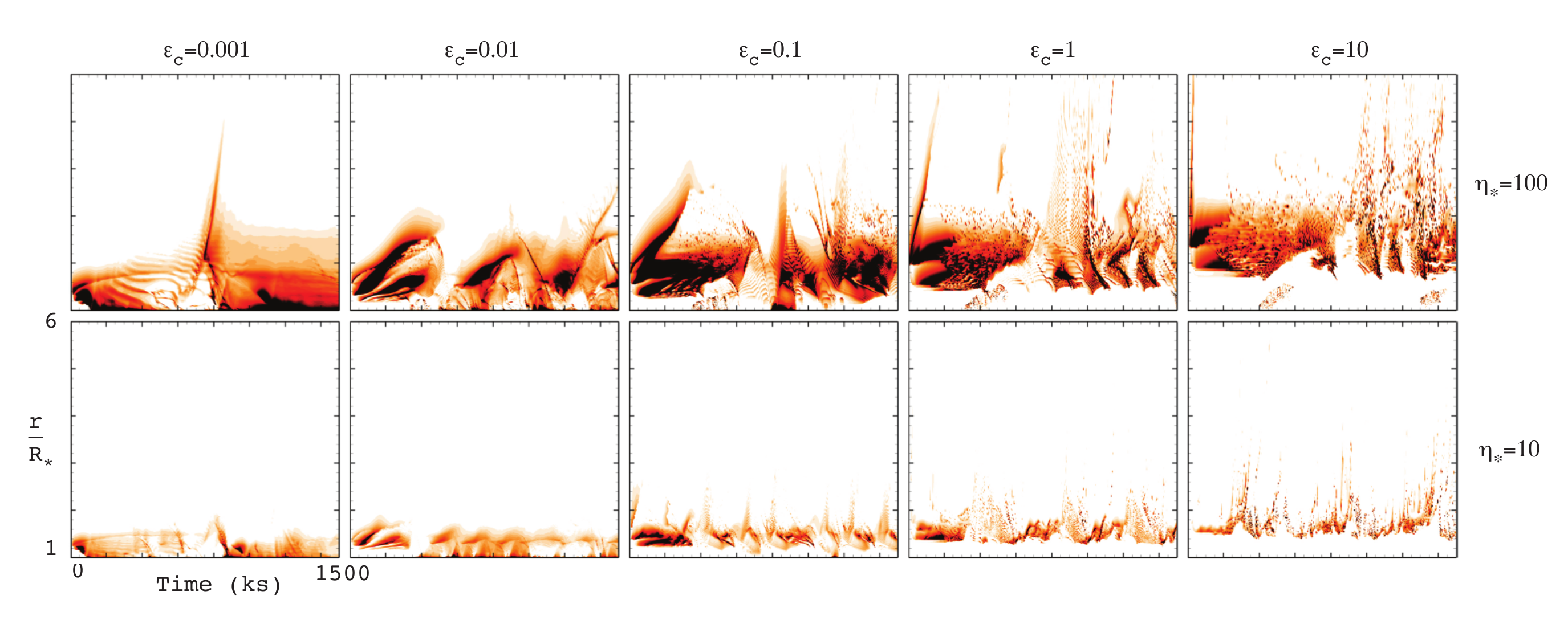}
\caption{
Mosaic of the radius and time variation of latitudinal- and energy-integrated X-ray emission above a threshold $E_x =$\,0.3\,keV for models with IC cooling and $\eta_\ast$=100 (upper row) or $\eta_\ast$=10 (lower row), with columns representing the 5 values of cooling efficiency $\epsilon_c$, ranging from $10^{-3}$ (left) to $10^1$ (right).
Between the $\eta_\ast$=10 vs.\ 100 models, the relative color strength reflects the relative X-ray luminosity. 
Within each $\eta_\ast$ row, the emission is scaled by the total $L_x$ for each $\epsilon_c$, and plotted on a common, linear color scale.
For  decreasing $\epsilon_c$ the decrease in the lower boundary radius for X-ray emission reflects the stronger shock retreat,  while the higher upper radial extent of X-rays in the $\eta_\ast=100$ vs. 10 models reflects the larger Alfv\'{e}n radius $R_A$.
}
\label{fig:mosaic}
\end{center}
\end{figure*}

\subsection{Results}
\label{sec:paramresults}

This limited effect of IC cooling is demonstrated clearly by the plots in figure \ref{fig:lxe-rad-vs-ic} of time-averaged X-ray spectra. 
The thick line curves and the regular thickness curves compare directly models with and without IC cooling, for the full set of 5 cooling efficiencies ranging from high ($\epsilon_c = 10$; blue curves at top) to low 
 ($\epsilon = 10^{-3}$; red curves at bottom), and 
 for confinement parameters $\eta_\ast = 100$ (left) and $\eta_\ast = 10$ (right). 
 The principal effect is to modestly reduce the high-energy emission for all cases, leading to generally softer X-ray spectra.
The strong cubic increase of radiative cooling with shock speed (eqn.\ (\ref{eq:trad})) means the radiative cooling is inefficient in the strongest shocks, but the addition of IC cooling, which is independent of shock speed, can still effectively cool such strong shocks, and thus reduce the hard X-ray emission they produce.
This effect of IC cooling in dissipating strong shocks, and so reducing and softening the X-ray emission, follows qualitatively the trends predicted by \citet{White95} in the context of colliding stellar winds.
But for X-rays from MCWS we see here that the overall importance of such effects is quite limited, and that to a reasonable approximation {\em one can largely ignore IC effects for modeling X-rays from magnetic stars.}

Figure \ref{fig:lxhrvseps} shows how changes in the cooling efficiency 
 $\epsilon_c$ affect both the total X-ray luminosity $L_x$ above some threshold $E_x = 0.3$\,keV (left) and the hardness ratio (H-S)/(H+S) (right), where H represents hard X-rays from  1 to 10 \,keV and S represents soft X-rays between 0.3 and 1\,keV.
The main trends are that lower efficiency (and so lower mass loss rate) lead toward lower luminosity and lower hardness.
The similarity between models with and without IC cooling (shown respectively by thick vs. thin curves) again illustrates the limited importance of IC cooling, except for the tendency toward somewhat softer spectrum in the high-mass-loss  radiative-shock limit, vs.\ somewhat harder spectra in the low-mass-loss, shock-retreat limit.

\subsection{Mosaic of radius-time plots for X-ray emission}
\label{sec:mosaic}

Finally, to gain insight on how this general shock-retreat scaling is maintained within the complex, time-dependent patterns of shock formation, cooling, and infall that occurs in the full MHD simulations, let us examine again the time and radius variation of the latitudinally integrated X-ray emission that was introduced in the right panel of figure \ref{fig:dxemdr}. 

Figure \ref{fig:mosaic} shows a mosaic of analogous time-radius plots of X-ray emission for various values of the cooling efficiency $\epsilon_c$ (in columns) and for the two magnetic confinement cases (top and bottom rows).
Within the complex variations from cycles of shock-formation and infall, note the broad patterns and trends for the characteristic height of X-ray emission.
Specifically,  in cases with lower efficiency, X-rays generally form at lower radii, reflecting the strong shock retreat.  The extent and strength of X-ray emission is greater in the model with stronger confinement, $\eta_\ast$=100.

\section{Analytic ``XADM'' scaling for $L_x$}
\label{sec:xadm}

\subsection{X-rays from confined loops with shock retreat}
\label{sec:analysis}

To help interpret these MHD results for X-rays, let us use a semi-analytic analysis to derive a generalized ``XADM"  scaling law for X-rays emitted from 
MCWS  in slowly rotating magnetic massive stars with dynamical magnetospheres.
For this we first note that, as shown in \citet{Owocki04c}, for a dipole magnetic field that intercepts the stellar surface at a co-latitude $\theta_\ast \equiv \arccos \mustar$, the local latitudinal variation of radial mass flux $\mdot$ (measured relative to the mass loss rate $\Mdot$ in the non-magnetic case)  
scales as\footnote{The normalization here accounts for equal contributions  from both north and south hemispheres, over an assumed restricted range, $0 < \mustar < 1$.}
\beq
\frac{d \mdot}{d \mustar} 
= \mu_B^2  = \frac{4 \mustar^2}{1+3 \mustar^2}
\, ,
\label{eq:dmddmu}
\eeq
where $\mu_B$ is the radial projection cosine of the local surface field, and the second equality applies to a standard dipole.
The maximum radius $r_m$ of the overlying dipole loop line occurs at the magnetic equator $\mu=0$, given in terms of the stellar radius $\Rstar$ by 
\beq
r_m = \frac{\Rstar}{1 - \mustar^2 }
\, .
\label{eq:rcdef}
\eeq
In terms of the total kinetic energy of the non-magnetized wind $L_{kin}=\Mdot \vinf^2/2$,
the associated latitudinal distribution of shock-dissipated energy can be written in terms of the scaled shock speed $w_s$,
\beq
\frac{dK_s}{d \mustar} =  \frac{d \mdot}{d \mustar} w_s^2 
=  \frac{4 \mustar^{2+4\beta}}{1+3 \mustar^2} \, \left ( \frac{w_s}{w_m} \right )^2 
\, .
\label{eq:dkmdmu}
\eeq

Following the analysis in \S 2.5 of Kee et al.\ 2014, we can write the fraction of this energy emitted as X-rays above a threshold energy $E_x$ as
\beq
f_x (T_s,E_x) = \int_0^{T_s} \frac{\Lambar (T,E_x)}{\Lambda(T)} \, \frac{dT}{T_s}
\, ,
\label{eq:fxdef}
\eeq
where the post-shock temperature $T_s = w_s^2 T_\infty$, with $T_\infty$ given by eqn.\ (\ref{eq:tsdef}) for $v_w = \vinf$.
Using the analysis in Appendix \ref{sec:appendix}, this can be approximated by
\beqa
f_x (T_s,E_x)  &\approx& \int_0^{T_s} e^{-E_x/kT} \frac{dT}{T_s} 
\label{eq:fxanal0}
\\
&=&
 e^{-E_x/kT_s} + \frac{E_x}{kT_s} \, 
% \Gamma [0,T_x/T_s]
 {\rm E_i} (-E_x/kT_s )
%  {\rm E_i} \left(- \frac{T_x}{T_s} \right  )
\, ,
\label{eq:fxanal}
\eeqa
where 
${\rm E_i}$ is the exponential integral.
The maximum shock temperature occurs for shocks at the full wind terminal speed, given by eqn.\ (\ref{eq:tsdef}) as 
%$T_\infty = 14\, V_8^2$\,MK.
$kT_\infty = 1.2\, V_8^2$\,keV.
%(eqn.\ \ref{eq:tsdef})
Thus if we define the X-ray energy ratio,
\beq
\epsilon_{xs} \equiv \frac{E_x}{kT_\infty} = \frac{E_x}{1.2 {\rm keV} \, V_8^2} 
\, ,
\label{eq:tauxi}
\eeq
then the variation of X-ray fraction $f_x$ depends on the reduced shock speed through $E_x/kT_s = \epsilon_{xs}/w_s^2$.

For a magnetosphere with closed loops extending over co-latitudes with $0 < \mu_\ast \le  \mu_c$, the ratio of total X-ray luminosity to wind kinetic energy is thus given by the integral,
\beq
\frac{L_x}{L_{kin}} 
= \int_0^{\mu_c} \frac{4 \mustar^{2+4\beta}}{1+3 \mustar^2}   \,  \left ( \frac{w_s}{w_m} \right )^2 \, f_x(T_s,E_x)  \, d \mustar  
 \, ,
\label{eq:lxlkin}
\eeq
where this latitudinal extent can be written in terms of a  maximum loop closure radius $r_c$, 
\beq
\mu_c  \equiv \sqrt{ 1 - \Rstar/r_c }
\, .
\label{eq:mucdef}
\eeq
Equations (9) and (10) of \citet{Uddoula08} give this closure radius  in terms of the magnetic confinement parameter,
\beq
\frac{r_c}{\Rstar} \approx 
%1 + 0.7 \left [ (\eta_\ast + 1/4)^{1/4} - (1/4)^{1/4} \right ]
0.5 + 0.7 (\eta_\ast + 1/4)^{1/4} 
\, .
\label{eq:rcetas}
\eeq

\begin{figure}
\begin{center}
\vfill
\includegraphics[scale=0.65]{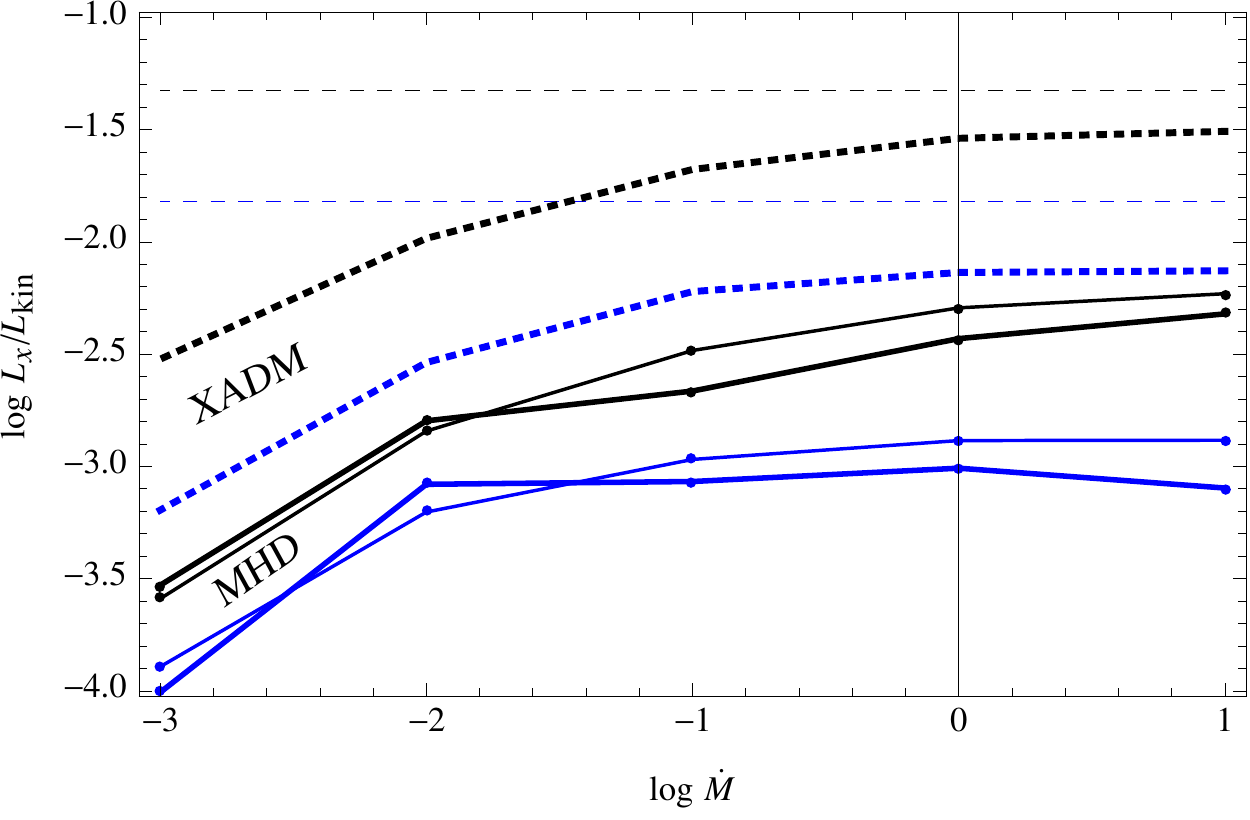}
\caption{The ratio of total X-ray luminosity $L_x$  from MCWS to the kinetic energy $L_{kin} = \Mdot \vinf^2/2$ in the non-magnetized wind, plotted vs.\  mass loss rate $\Mdot$ 
(scaled in terms of 
the standard model with $\Mdot = 3.1 \times 10^{-6} \Msun$/yr),
for cases $\eta_\ast$=\,10 (blue) and 100 (black).
The heavy and light solid curves are time-averaged values for numerical MHD simulations with and without IC cooling,  while the dotted curves are for the XADM analytic scaling in eqn.\ (\ref{eq:lxlkin}), 
using the dipole-shock-retreat analysis of Appendix \ref{sec:nonsph}.
The horizontal dashed lines give the upper limits for energy dissipated in MCWS, obtained from eqn.\ (\ref{eq:ksdef}) by assuming $ w_s/w_m = f_x = 1$ in the analysis leading to eqn.\ (\ref{eq:lxlkin}).
The infall and variability of the full MHD simulations makes the X-ray emission about a factor 5 lower than in the idealized, steady-state XADM model.
}
\label{fig:lxlkin}
\end{center}
\end{figure}

For context, a simple upper limit to the X-ray ratio (\ref{eq:lxlkin}) can be written for the case of strong radiative shocks with $w_s/w_m=f_x = 1$, for which the total dissipated kinetic energy in the magnetosphere is
\beq
K_s (\eta_\ast) =  \int_0^{\mu_c} \frac{dK_s}{d \mustar}    \, d \mustar  
=  \int_0^{\mu_c} \frac{4 \mustar^{2+4\beta}}{1+3 \mustar^2}  \, d \mustar  
\approx C_c  \frac{\mu_c^{3+4\beta}}{3+4\beta}
 \, .
\label{eq:ksdef}
\eeq
The last approximation ignores the denominator term in the integrand, with $C_c$ an order-unity correction; the resulting power-law form illustrates the strong dependence on closure latitude, i.e. as $\mu_c^7$ for a standard $\beta=1$ velocity law.
The full integration can be evaluated analytically with hypergeometric functions.
For $\beta=1$, the limit of arbitrarily strong confinement $\eta_\ast \rightarrow \infty$, for which $\mu_c \rightarrow 1$, gives $K_c =$\,0.177, implying then that even in this extreme limit  less than 18\% of wind kinetic energy is dissipated in MCWS.
For the MHD confinement cases $\ \eta_\ast =$ 10 and 100, the corresponding percentages ($100 K_c$\%) are 1.5\% and 4.7\% (see horizontal dashed lines in figure \ref{fig:lxlkin}).

\subsection{Comparison between analytic and numerical MHD scalings}

More generally, computation of the X-ray ratio (\ref{eq:lxlkin}) requires evaluation of the scaled shock speed $w_s$ after accounting for shock retreat, as given by the analysis in \S \ref{sec:shockretreat}, 
extended in Appendix \ref{sec:nonsph} to account for the dipole loop geometry.
Using standard root finding, one can readily solve 
(\ref{eq:wsdip})
for $w_s$  for any given values of the cooling efficiency $\chi_\infty$ [from eqn.\  (\ref{eq:chiinf})], and loop apex speed $w_m$.

For a given X-ray energy parameter $\epsilon_{xs}$, this then also gives the X-ray energy ratio, $E_x/kT_s = \epsilon_{xs}/w_s^2$, and so the X-ray fraction $f_x$ through (\ref{eq:fxanal}).
Since $w_m = \mustar^2$, $w_s$ and thus $f_x$ can be readily evaluated in carrying out the $\mustar$ integral (\ref{eq:lxlkin}), with
the integral upper bound $\mu_c$ depending on $\eta_\ast$ through eqns. (\ref{eq:mucdef}) and (\ref{eq:rcetas}).

The upshot is that the value of $L_x/L_{kin}$ is entirely set by the 3 dimensionless parameters $\eta_\ast$, $\chi_\infty$, and $\epsilon_{xs}$.

Evaluating  (\ref{eq:lxlkin}) in this way, figure \ref{fig:lxlkin}  plots this semi-analytic scaling for $L_x/L_{kin}$ vs.\ $\Mdot$ for $\eta_\ast =$ 10 and 100
(lower and upper dotted curves);
the thick and thin solid curves show analogous  time-averaged X-ray emission for MHD simulations with and without IC cooling.

The XADM scaling follows a very similar trend to the full MHD simulation results, but is about a factor 5 higher.
Compared to the idealized steady-state emission of the analytic XADM model, the numerical simulations show an extensive time variability with repeated intervals of infall of cooled, trapped material, and it appears this lowers the overall efficiency of X-ray emission to about 20\% of the idealized XADM prediction.

\begin{figure}
\includegraphics[scale=0.65]{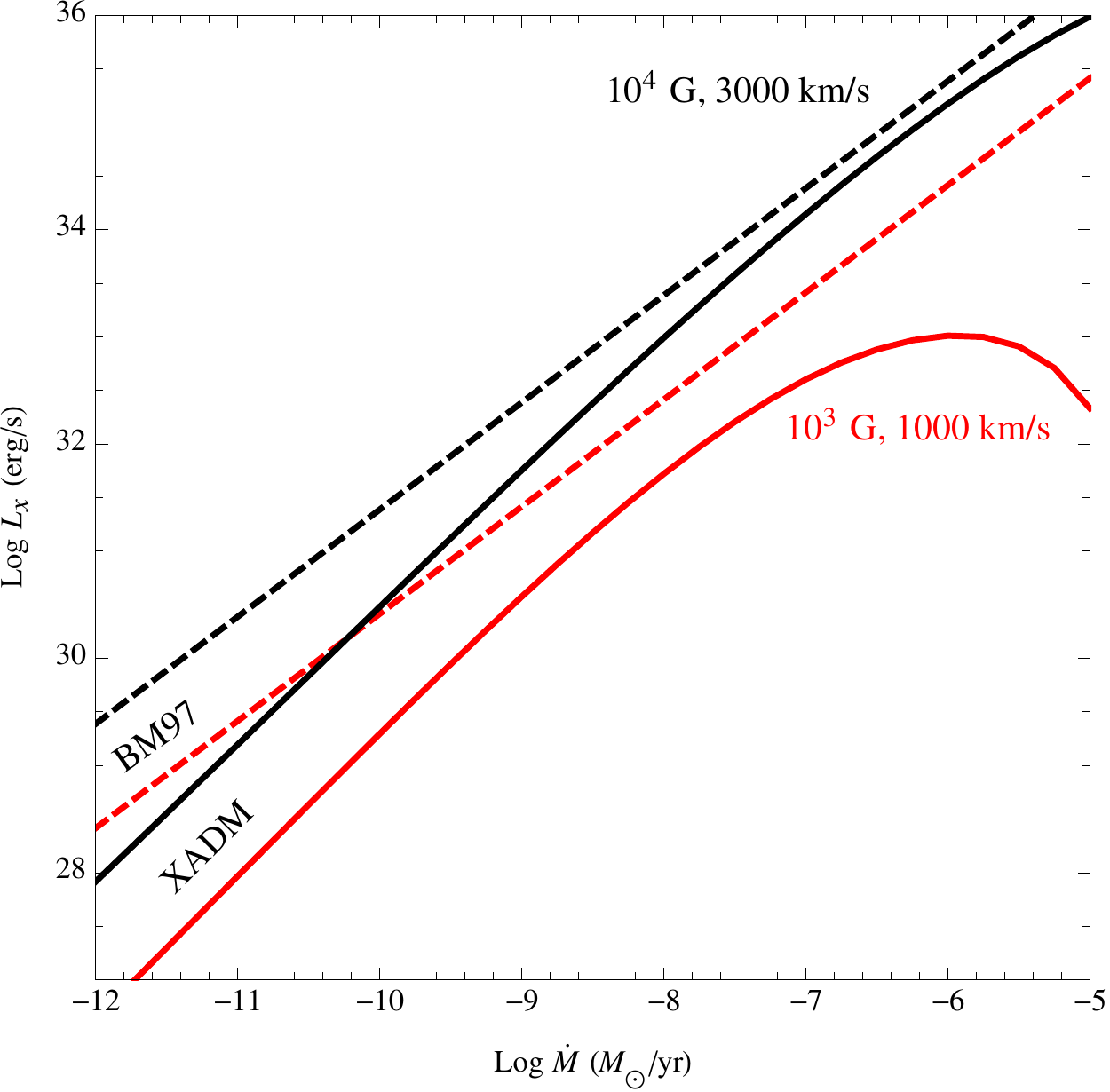}
\caption{Log-log plot of X-ray luminosity from XADM scaling (\ref{eq:lxlkin}) vs.\ mass loss rate for a ``high'' case with large field and fast  wind speed ($B_p=10^4$\,G, $\vinf =$\,3000\,km/s; black curve),  and a ``low" case with smaller field and slower speed ($B_p=10^3$\,G, $\vinf =$\,1000\,km/s; red curve).
The dashed lines compare the corresponding pure power-law scaling suggested by \citet{Babel97} (see their eqns.\ 10 and 11, with $\delta=1$ and $\epsilon=1$).
}
\label{fig:lxvsmdot}
\end{figure}

\subsection{Scaling recipe for interpreting observed X-rays}

Notwithstanding this overall difference in X-ray efficiency, the good general agreement in the trends for the MHD and XADM  encourages application of this semi-analytic XADM scaling to analyze the X-ray emission from magnetospheres with a broader range of magnetic and stellar properties than considered in the detailed MHD simulations here.

The model X-ray luminosity $L_x$ can be obtained by simple numerical evaluation of the integral formula (\ref{eq:lxlkin}), 
using the auxiliary eqns.\  (\ref{eq:wsdip}), (\ref{eq:fxanal}),  (\ref{eq:mucdef}),  and(\ref{eq:rcetas}), and then multiplying this by the wind kinetic energy $L_{kin} = \Mdot \vinf^2/2$.

As noted, this integral evaluation depends on three dimensionless parameters, namely: the magnetic confinement parameter $\eta_\ast$ [defined in eqn.\ (\ref{eq:etasdef})]; the cooling parameter $\chi_\infty$ [defined in eqn.\ (\ref{eq:chiinf})]; and ratio of  X-ray energy to terminal speed shock energy $\epsilon_{xs}$ [defined in eqn.\ (\ref{eq:tauxi})].

These in turn just depend on 
four  physical parameters: 
the surface field strength $B$, 
 the stellar radius $\Rstar$, 
 and the mass loss rate $\Mdot$ and terminal speed $\vinf$ that would occur in a {\em non}-magnetic stellar wind for the inferred stellar parameters (i.e., luminosity $L$ and mass $M$).
 
 The upshot is that for any slowly rotating magnetic massive star with an observed large-scale dipole field, estimating the stellar radius and mass loss parameters allows one to use this semi-analytic scaling (\ref{eq:lxlkin}) to predict an X-ray luminosity from MCWS, and then compare this against observed values to test the applicability this MCWS paradigm.
 
Figure \ref{fig:lxvsmdot}  plots $L_x$ vs.\ $\Mdot$ (on a log-log scale) for two cases intended to roughly bracket the range in X-ray emission, namely a ``high'' case with large field and fast  wind speed ($B_p=10^4$\,G, $\vinf =$\,3000\,km/s; black curve) and a ``low" case with smaller field and slower speed ($B_p=10^3$\,G, $\vinf =$\,1000\,km/s; red curve).
The dashed lines compare the pure power-law scaling suggested by \citet{Babel97},
\beq
L_x = 2.6 \times 10^{30} {\rm erg \, s^{-1}} \, 
\Mdot_{-10} \, V_8 \, B_{3}^{0.4} 
\, ,
\label{eq:lxbm}
\eeq
where $\Mdot_{-10} \equiv \Mdot/(10^{-10} \, \Msun$/yr) and $B_3 \equiv B_p/10^3$\,G.
Remarkably, the two scalings are  quite comparable at moderate mass loss rate.
But at low $\Mdot$ the shock retreat causes the semi-analytic X-rays to drop more steeply than the linear $\Mdot$ scaling assumed by BM97a. 
Moreover, at high $\Mdot$, the reduction of magnetic confinement (to $\eta_\ast $ approaching unity) for the case with lower field ($B_p=$1000\,G) causes a flattening and even turnover in $L_x$, again making this fall well below the linear scaling for BM97a.

This demonstrates quite clearly the importance of both shock retreat and magnetic confinement in setting the mass-loss scaling of X-ray luminosity from MCWS.
In applying this XADM scaling to interpreting X-ray observations, it would be appropriate to reduce the predicted $L_x$ by an efficiency factor $\sim 0.2$ to account for lower average emission from dynamical models with infall of trapped material.

\section{Summary \& Future Work}

This paper uses MHD simulations to examine the effects of radiative and inverse-Compton (IC) cooling on X-ray emission from magnetically confined wind shocks (MCWS) in the dynamical magnetospheres (DM) that arise in slowly rotating magnetic massive stars with radiatively driven (CAK) stellar winds.
The key results can be summarized as follows:
\begin{itemize}
\item
The scaling of IC cooling with luminosity and radiative cooling with mass loss rate suggests that for CAK winds with $\Mdot \sim L^{1.7}$, IC cooling should become relatively more important for lower luminosity stars. 
However, because the sense of the trends is similar, including IC cooling  has a quite modest overall effect  on the broad scaling of X-ray emission.
\item 
For the two fixed values of magnetic confinement ($\eta_\ast$=10, 100) used in MHD simulations here, the reduced efficiency of radiative cooling from a lower mass loss rate causes a shock retreat to lower speed wind, leading to weaker shocks. This lowers and softens the X-ray emission, making the $\Mdot$ dependence of $L_x$ steeper than the linear scaling seen at higher $\Mdot$ without shock retreat.
\item
These overall scalings of time-averaged X-rays in the numerical MHD simulations are well matched by the $L_x$ computed from a semi-analytic ``XADM'' model that accounts for both shock retreat and magnetic confinement within the context of steady feeding of the DM by a  CAK wind with field-adjusted mass flux. 
 However, the values of $L_x$ are about a factor 5 lower in the MHD models, mostly likely reflecting an overall inefficiency of X-ray emission from the repeated episodes of dynamical infall.
\item 
Comparison with the previous power-law scaling ($L_x \sim \Mdot \vinf  B^{0.4}$)  suggested by BM97a shows a general agreement with XADM at intermediate $\Mdot$. But the XADM $L_x$ drops well below the power-law scaling at both low $\Mdot$ (due to shock retreat) and high $\Mdot$ (due to weakened magnetic confinement).
\item
The XADM reproduction of  trends in MHD X-rays encourages application of this XADM scaling, with a factor 0.2 efficiency reduction, toward interpreting X-ray observations of slowly rotating magnetic massive stars with a broader range of field strength and wind parameters than considered in the MHD simulations here.
\end{itemize}

Within this theoretical framework, one focus of our future work will be to apply these results toward interpreting X-ray observations for the subset of  confirmed magnetic massive stars \citep{Petit13} with available X-ray data from {\em Chandra} or {\em XMM-Newton}, with initial emphasis on slowly rotating O and B stars. 
(See \citet{Naze14}.)
%(See Naz\'e et al. 2014.)
To facilitate analysis of the moderately fast rotating B-stars with centrifugal magnetospheres (CM), we also plan an extension of the present simulation study to examine the potential effects of rotation on the X-ray emission.

\section*{Acknowledgments}
Support for this work was provided by NASA through Chandra Award numbers TM4-15001A, TM4-1500B and TM4-15001C issued by the 
Chandra X-ray Observatory Center which is operated by the Smithsonian Astrophysical Observatory for and behalf of NASA under contract
NAS8-03060. This work was also carried out with partial support by NASA ATP Grants  NNX11AC40G  
and NNX08AT36H S04, respectively to the University of Delaware and the University of Wisconsin.
D.H.C. acknowledges support from NASA
ADAP grant NNX11AD26G and NASA {\it Chandra} grant AR2-13001A to Swarthmore College.
We thank M.~Gagn\'{e}  and G.~Wade for many helpful discussions.
We especially  thank C.~Russell for providing us tabulations of the radiative emission function $\Lambda (E,T)$ through his application of the APEC model in XSPEC.

\bibliographystyle{mn2e}
\bibliography{OwockiS}

\appendix

\section{Radiative emission function}
\label{sec:appendix}

\begin{figure}
\begin{center}
\vfill
\includegraphics[scale=0.50]{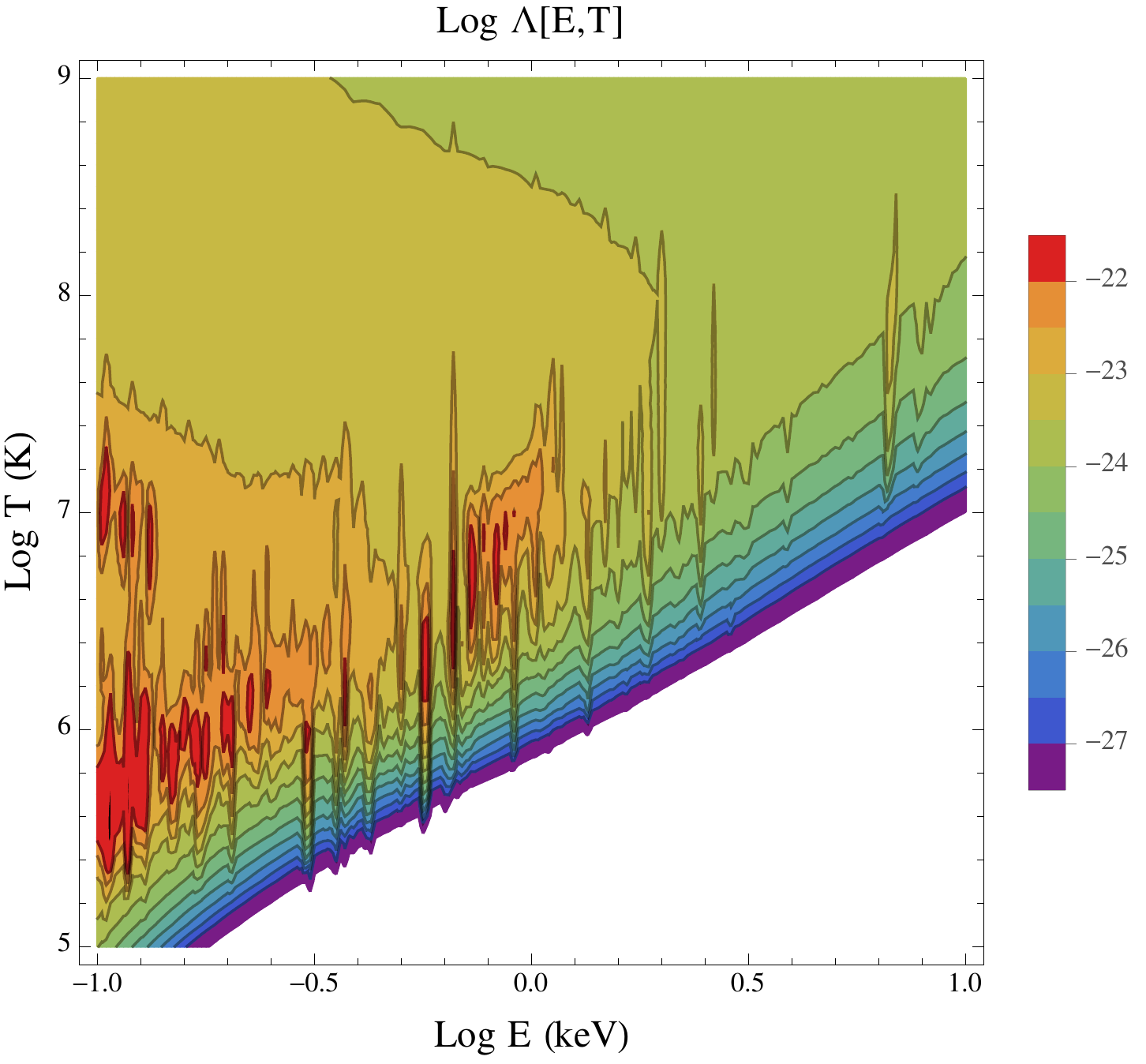}
\caption{
Log of emission function
$ \Lambda (E,T)$  (erg\,cm$^3$/s) in logarithmic energy bins, plotted vs. log of photon energy (in keV) and log temperature (in K).
}
\label{fig:lamplts}
\end{center}
\end{figure}

\begin{figure*}
\begin{center}
\vfill
\includegraphics[scale=0.65]{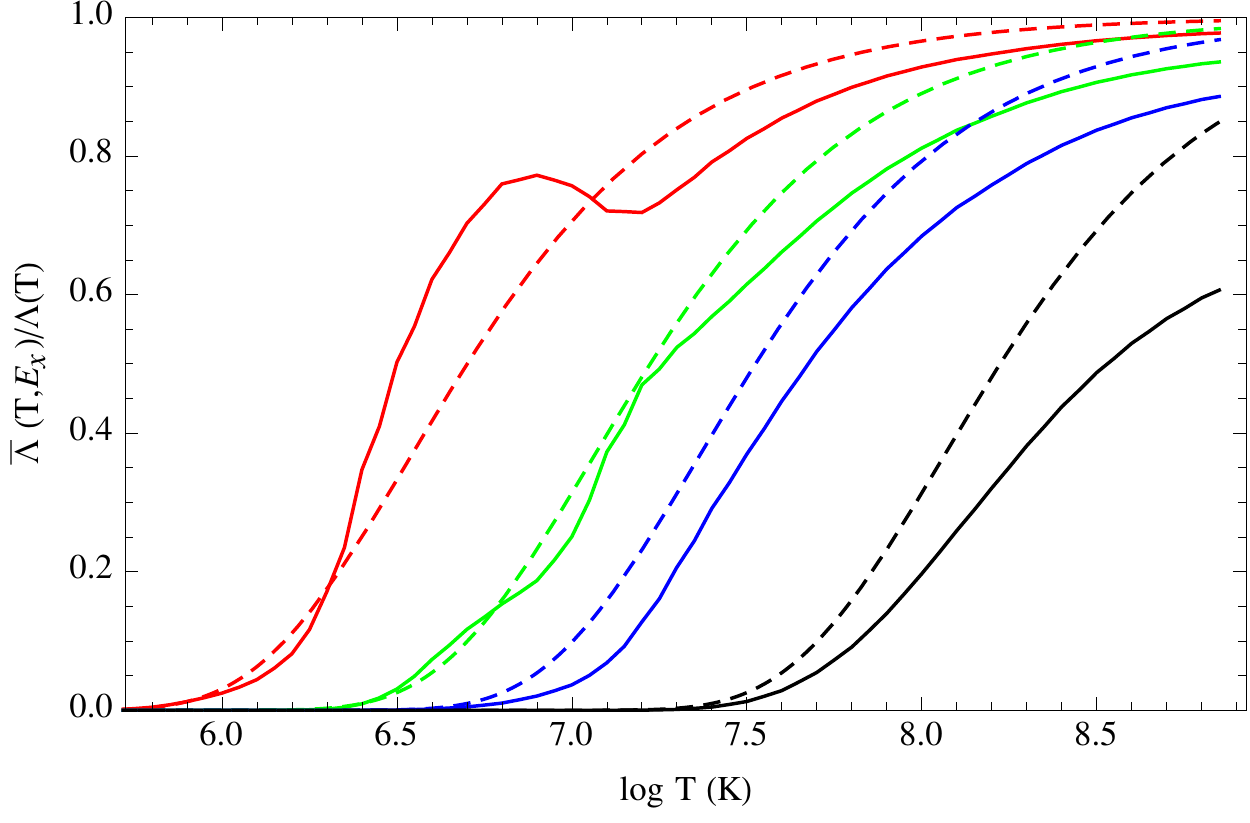}
~~
\includegraphics[scale=0.65]{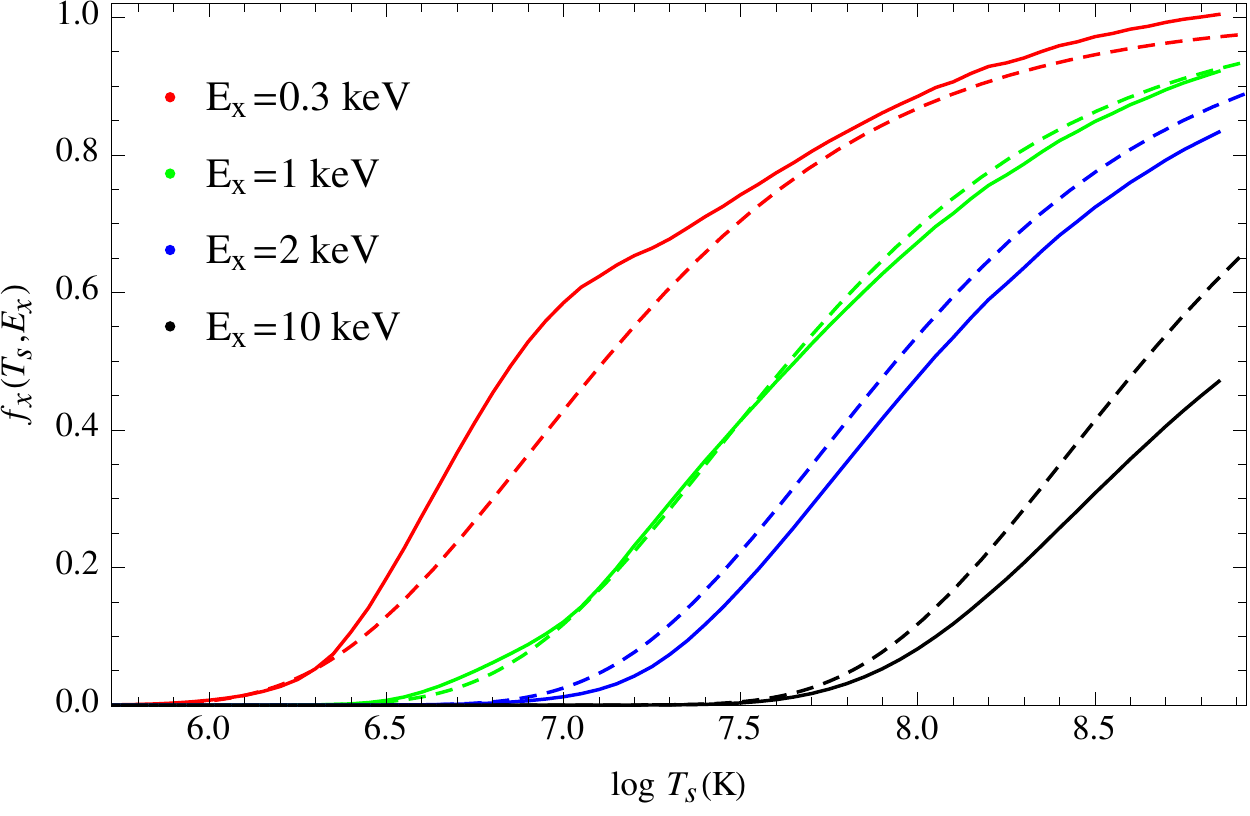}
\caption{{\em Left:} Temperature variation of the ratio ${\bar \Lambda} (T,E_x)/\Lambda(T)$ for $E_x =$\,0.3, 1, 2 and 10\,keV; the dashed curves compare the simple Boltzmann function fits used in the integrand of eqn.\ (\ref{eq:fxanal0}).
{\em Right:}  Associated numerical evaluation of the shock temperature integral (\ref{eq:fxdef}) for the same four X-ray threshold energies;
the dashed curves compare the analytic function in eqn.\ (\ref{eq:fxanal}).
}
\label{fig:boltz}
\end{center}
\end{figure*}

X-ray spectra in this paper are computed from the energy- and temperature-dependent emission function $\Lambda(E,T)$, as tabulated from the APEC model 
\citep{Smith01, Foster12} in the XSPEC plasma emission code
\citep{Arnaud96}.
Figure \ref{fig:lamplts} gives a color plot of $\log \Lambda (E,T)$ vs. the log of energy  and temperature.

Integration of this emission function from an energy threshold $E_x$ gives the cumulative function ${\bar \Lambda}(T,E_x)$, with the total cooling function approximated by the value for the lowest tabulated energy, $\Lambda(T) \approx {\bar \Lambda} (T,E_{min})$, where here for tables used $E_{min} = 0.01$\,keV.
The left panel of figure \ref{fig:boltz} plots the temperature variation of the ratio ${\bar \Lambda} (T,E_x)/\Lambda(T)$ for $E_x =$\,0.3, 1, 2 and 10\,keV; the dashed curves show the simple Boltzmann function fits used in the integrand of eqn.\ (\ref{eq:fxanal0}).
The right panel of  plots associated numerical evaluation of the shock temperature integral (\ref{eq:fxdef}) for the same four X-ray threshold energies;
the dashed curves compare the analytic function in eqn.\ (\ref{eq:fxanal}).

\section{Shock retreat along a dipole loop}
\label{sec:nonsph}

Let us now generalize the simplified spherical shock-retreat model of \S \ref{sec:shockretreat} to account for the flow geometry along a dipole loop.
For a flow tube along a coordinate $s$ with cross sectional area $A$, we can write eqn.\ (\ref{eq:dtdrshock}) in the generalized form
\beqa
\frac{T^2 dT}{T_s^3} 
&=& - \frac{2}{5} \, \frac{\mubar \Lambda_m}{kT_s} \, \frac{\rho T^2}{v T_s^2} \, ds
\\
&=& - \frac{2}{5} \, \frac{\mubar \Lambda_m}{kT_s} \, \frac{\rho^2 T^2}{\Mdot T_s^2} \, A \, ds
\\
&=& - \frac{32}{5} \, \frac{\mubar \Lambda_m}{kT_s} \, \frac{\rho_{ws}^2 T_s^2}{\Mdot T_s^2} \, A \, ds
\\
&=& - \frac{512}{15} \, \frac{\mubar \Lambda_m}{v_s^2} \, \frac{\Mdot}{v_s^2 A_s^2 } \, A \, ds
\\
&=& - \frac{1}{\chi_\infty} \, \frac{\mdot}{w_s^4} \, \frac{A_\ast}{A_s^2 \Rstar} \, A \, ds
\, ,
\eeqa
where $\mdot$ allows for a mass loss weighting for a given flow tube, defined as a fraction of the spherical mass loss $\Mdot$ used in the definition of $\chi_\infty$.
Integration from the shock radius $r_s$ gives the temperature variation,
\beq
1- \left ( \frac{T}{T_s} \right )^3 = \frac{3}{\chi_\infty} \, \frac{\mdot}{w_s^4} \, \frac{A_\ast}{A_s^2 \Rstar} \, \int_{r_s}^r A \, ds 
\, .
\eeq
Setting the apex temperature $T(r_m) = 0$ then allows us to cast a general implicit equation for the shock radius $r_s$,
\beq
g(r_s/r_m) \equiv  \int_{r_s/r_m}^1\frac{A}{A_m} \, \frac{ds}{r_m} =   \frac{\chi_\infty}{3} \, \frac{w_s^4}{\mdot} \, \frac{A_s^2}{A_\ast A_m} \, \frac{\Rstar}{r_m}
\, .
\eeq
For the spherical case with $ds=dr$, $A \sim r^2$ and $\mdot =1$,
\beq
1-\left ( \frac{r_s}{r_m} \right )^3 
%= \chi_\infty w_s^4 \frac{r_s}{\Rstar} \, \left ( \frac{r_s}{r_m} \right )^3
= \chi_\infty  \, \left ( \frac{w_s r_s}{r_m} \right )^4 \, \frac{r_m}{\Rstar}
\, ,
\label{eq:wssph}
\eeq
which is equivalent to (\ref{eq:implicitrs}).

For flow along a dipole magnetic field line with base co-latitude set by $\mustar$, we have 
%$A \sim 1/B$, and 
\beq
\frac{A}{A_\ast} = \frac{B_\ast}{B} = \left ( \frac{r}{\Rstar} \right )^3  \frac{\sqrt{1+3 \mustar^2}}{\sqrt{1+3 \mu^2}} 
\, ,
\eeq
where $r = (1-\mu^2) r_m$, and $r_m = \Rstar/(1-\mustar^2)$.
Also the differential along the field line coordinate can be written
\beq
ds = \frac{r d \theta}{{\hat B}_\theta} = - \frac{r \sqrt{1+3\mu^2}}{1-\mu^2} \, d\mu
\, ,
\eeq
where ${\hat B}_\theta$ is the unit field projection in the $\theta$ (latitudinal) direction.
Thus
\beqa
g(r_s/r_m)
&=&  \int_{r_s/r_m}^1\frac{B_m}{B} \, \frac{ds}{r_m} 
%\\
%&=&  \int_0^{\mu_s} \left ( \frac{r}{r_m}  \right )^3 \frac{1}{\sqrt{1+ 3 \mu^2}} \, \frac{r}{r_m} \, \frac{\sqrt{1+3\mu^2}}{1-\mu^2} \, d\mu
\\
&=&  \int_0^{\mu_s} \left ( \frac{r}{r_m}  \right )^4 \, \frac{d\mu}{1-\mu^2} 
\\
&=&  \int_0^{\mu_s}  ( 1-\mu^2 )^3 \, d\mu
\\
&=& 
\mu_s - \mu_s^3 + \frac{3 \mu_s^5}{5} - \frac{\mu_s^7}{7}
\, ,
\label{eq:gdip}
\eeqa
where $\mu_s \equiv \sqrt{1-r_s/r_m}$.

As done for spherical shock retreat in \S \ref{sec:shockretreat}, for a given $\chi_\infty$ and $w_m = \mustar^2$, 
we can now use this analytic formula (\ref{eq:gdip}) for $g$ to solve for $r_s$ (and, for a $\beta=1$ law, for $w_s = 1 - \Rstar/r_s$), 
through the implicit equations,
%\beqa
%g(r_s/r_m) 
%&=&   \frac{\chi_\infty}{3}  \, \frac{w_s^4}{\mdot} \, \frac{B_\ast B_m}{B_s^2} \, \frac{\Rstar}{r_m}
%\\
%\mu_s - \mu_s^3 + \frac{3 \mu_s^5}{5} - \frac{\mu_s^7}{7}
%&=&  \chi_\infty \, \left( \frac{w_s r_s}{r_m} \right )^4  \left( \frac{r_s}{\Rstar} \right )^2 \,  \frac{1}{6\mustar} 
% \, \frac{1+3\mustar^2}{1+3\mu_s^2} 
%\label{eq:wsdip}
%\eeqa
\beq
g(r_s/r_m) 
=   \frac{\chi_\infty}{3}  \, \frac{w_s^4}{\mdot} \, \frac{B_\ast B_m}{B_s^2} \, \frac{\Rstar}{r_m}
\label{eq:wsdip0}
\eeq
\beq
\mu_s - \mu_s^3 + \frac{3 \mu_s^5}{5} - \frac{\mu_s^7}{7}
=   \frac{\chi_\infty}{6\mustar} 
 \, \frac{1+3\mustar^2}{1+3\mu_s^2} \,  \left( \frac{w_s r_s}{r_m} \right )^4  \left( \frac{r_s}{\Rstar} \right )^2 
 \, ,
\label{eq:wsdip}
\eeq
where the second equality uses a weighting  $\mdot = 2 \mustar/\sqrt{1+3\mustar^2}$ for the mass flux {\em along} a field line with base latitude set by $\mustar$ \citep{Owocki04c}.

The solid curves in figure \ref{fig:wsvschi} plot the variation of $w_s$ vs.\ $\chi_\infty$ for various loop lines with scaled apex speed $w_m$ from 0.1 to 0.9. 
The dashed curves compare results for the simplified spherical shock-retreat example of \S \ref{sec:shockretreat}.
The differences only become significant for large $\chi_\infty$ (low $\Mdot$), but for completeness we use this full dipole shock retreat in the XADM scaling analysis of \S \ref{sec:xadm}.

\end{document}